\documentclass[12pt,a4paper]{article}
\usepackage[utf8]{inputenc}
\usepackage{amsmath}
\usepackage{amsfonts}
\usepackage{amssymb}
\usepackage{graphicx, datetime, abstract}
\usepackage{float}
\usepackage{fullpage}
\usepackage{caption}
\usepackage{subcaption}
\usepackage{todonotes}
\usepackage{setspace}
\usepackage{resizegather}
\pdfoutput=1

\begin{document}
%	\doublespacing
	\title{
			\begin{flushright}\ \vskip -2cm {\small{\em DCPT-15/47}}\end{flushright}
		Baby Skyrmions in AdS}
	\author{Matthew Elliot-Ripley$^*$ and Thomas Winyard$^\dagger$\\[10pt]
		{\em \normalsize Department of Mathematical Sciences, }\\{\em \normalsize Durham University, Durham, DH1 3LE, U.K.}\\[10pt]
		{\normalsize $^*$m.k.i.d.elliot-ripley@durham.ac.uk\quad $^\dagger$t.s.winyard@durham.ac.uk}}
		\date{July 2015}
	\maketitle
	\vspace{15pt}
	\begin{abstract}
We study the baby Skyrme model in a pure AdS background without a mass term. The tail decays and scalings of massless radial solutions are demonstrated to take a similar form to those of the massive flat space model, with the AdS curvature playing a similar role to the flat space pion mass. We also numerically find minimal energy solutions for a range of higher topological charges and find that they form concentric ring-like solutions. Popcorn transitions (named in analogy with studies of toy models of holographic QCD) from an $n$ layer to an $n+1$-layer configuration are observed at topological charges $9$ and $27$ and further popcorn transitions for higher charges are predicted. Finally, a point-particle approximation for the model is derived and used to successfully predict the ring structures and popcorn transitions for higher charge solitons.
	\end{abstract}
	
	\newpage 

\clearpage
\graphicspath{{./AdSPics/}}
\section{Introduction}
The Skyrme model \cite{Skyrme:1961vq} is a non-linear theory of pions in $(3+1)$ dimensions, admitting soliton solutions called Skyrmions. It has been derived as a low-energy effective field theory of QCD in the large colour limit, and has also been used in holographic models such as the Sakai-Sugimoto model \cite{Sakai:2004cn, Sakai:2005yt} where Yang-Mills Chern-Simons instantons in a $(4+1)$-dimensional bulk spacetime are dual to (extended) Skyrmions on the boundary. A key feature of these models is that the bulk spacetime is AdS-like, possessing a conformal boundary and a finite, negative scalar curvature.

Solitons in pure Anti de-Sitter spacetimes are also of interest. It has been shown that Skyrmions with massless pions in hyperbolic space are closely related to Skyrmions with massive pions in Euclidean space \cite{Atiyah:2004nh, Winyard:2015ula}. Since constant time slices of AdS spacetimes are hyperbolic we may expect similar results in pure AdS. In addition, monopoles and monopole walls have been studied in AdS \cite{Bolognesi:2010nb, Sutcliffe:2011sr}, motivated as a magnetic version of holographic superconductors.

The baby Skyrme model \cite{Piette:1994ug} is a $(2+1)$ dimensional analogue of the Skyrme model. Baby Skyrmions have recently been used to study low-dimensional models of the Sakai-Sugimoto model in the context of dense QCD \cite{Bolognesi:2013jba, Elliot-Ripley:2015cma}. In these toy models a series of phase-transitions were observed in which infinite chains of solitons split into multiple layers with increasing density. These were dubbed \emph{popcorn transitions}, and the extra layers were found to be separated in the holographic direction.

Here we investigate the baby Skyrme model in a pure AdS background and study the resulting soliton and multi-soliton solutions. The low dimensionality of the model makes full numerical field computations viable, and we find that the curvature of the spacetime allows us to find soliton solutions even without a pion mass term. Multi-solitons beyond topological charge $B=3$ are found to take the form of ring-like structures, with popcorn-like phase transitions to multi-layered rings occurring as the topological charge increases. Inspired by methods used to study aloof baby Skyrmions \cite{Salmi:2014hsa} we will derive a point-particle approximation to study these phase transitions for higher topological charges.

\section{AdS spacetime in $(2+1)$ dimensions}
AdS is the maximally symmetric spacetime with Minkowskian signature and constant negative curvature, and in this paper we will be interested in the $(2+1)$-dimensional case. The metric, given in \emph{sausage coordinates}, can be written
\begin{equation}\label{metric}
ds^2 = -\left(\frac{1+r^2}{1-r^2}\right)^2dt ^2+\frac{4 L^2}{(1-r^2)^2}\left(dx^2 + dy^2\right)\, .
\end{equation}
Here, $L$ is the AdS radius (related to the cosmological constant $\Lambda$ via $\Lambda = -1/L^2$) and $r=\sqrt{x^2+y^2}\in[0,1)$ is a radial coordinate. These coordinates are useful numerically as they are global coordinates over a finite range. Later we will use that the geodesic distance between two points $\pmb{x}$ and $\pmb{y}$ in this spacetime is given by
\begin{equation}\label{sausagedist}
d(\pmb{x},\pmb{y}) = L\cosh^{-1}\!\left(1+\frac{2|\pmb{x}-\pmb{y}|^2}{(1-|\pmb{x}|^2)(1-|\pmb{y}|^2)}\right)\,.
\end{equation}

The following \emph{global coordinates} give another useful way of writing the AdS metric:
\begin{equation}\label{metricglob}
ds^2 = -\cosh^2\!\frac{\rho}{L}\, dt^2+d\rho^2 + L^2\sinh^2\!\frac{\rho}{L}\, d\theta^2\, ,
\end{equation}
where $\rho\ge0$. These coordinates are useful since the radial coordinate $\rho$ coincides with the geodesic distance from the origin in this model. The global and sausage coordinates are related by $r = \tanh{\frac{\rho}{2L}}$.
%
%Finally, we can write the spatial slices of this metric using a complex coordinate $z=x+iy$ as
%\begin{equation}\label{cxmetric}
%ds^2 = \frac{4L^2}{\left(1-|z|^2\right)^2}|dz|^2\,.
%\end{equation}
%In these coordinates the spatial part of the metric is invariant under M\"obius maps of the form
%\begin{equation}
%h(z) = e^{i\theta}\frac{z-a}{1-\bar{a}z}\,,
%\end{equation}
%for $\theta\in\mathbb{R}$ and $|a|\le 1$. Setting $\theta=0$ and converting back to $(x,y)\equiv\pmb{x}$ coordinates allows us to define a \emph{hyperbolic translation} in sausage coordinates that sends the origin to $\pmb{a}$ by
%\begin{equation}\label{ADStrans}
%\pmb{x}\mapsto \frac{(1-|\pmb{a}|^2)\pmb{x}+(1+2\pmb{x}\cdot\pmb{a}+|\pmb{x}|^2)\pmb{a}}{1+2\pmb{x}\cdot\pmb{a}+|\pmb{a}|^2|\pmb{x}|^2}\,.
%\end{equation}

Since constant time slices of AdS are hyperbolic space, it will be useful to define a \emph{hyperbolic translation} in sausage coordinates. A translation sending the origin to some point $\pmb{a}$ is given by
\begin{equation}\label{ADStrans}
\pmb{x}\mapsto \frac{(1-|\pmb{a}|^2)\pmb{x}+(1+2\pmb{x}\cdot\pmb{a}+|\pmb{x}|^2)\pmb{a}}{1+2\pmb{x}\cdot\pmb{a}+|\pmb{a}|^2|\pmb{x}|^2}\,.
\end{equation}
It should be noted that this is an isometry of only the constant time slices of the spacetime, rather than the full spacetime, since the time component of the metric is dependent on $r$.

As a final note, the Ricci scalar curvature of AdS can be calculated as $R=-6/L^2$. In the limit $L\to\infty$ this curvature vanishes and we recover flat space.

\section{The AdS baby Skyrme model}
We will be interested in studying soliton solutions of the baby Skyrme model, given by the action
\begin{align}\label{bsaction}
S_{BS} = -\frac{1}{2}\int\! \Big( \partial_\mu\pmb{\phi}\cdot\partial^\mu\pmb{\phi} + \frac{\kappa^2}{2}(\partial_\mu\pmb{\phi}\times&\partial_\nu\pmb{\phi})\cdot(\partial^\mu\pmb{\phi}\times\partial^\nu\pmb{\phi}) \nonumber \\
& + 2 m^2(1-\pmb{\phi}\cdot\pmb{n})\Big)\sqrt{-\det{g}}\, dx\, dy\, dt\,.
\end{align}
The field $\pmb{\phi} = (\phi_1 , \phi_2 , \phi_3)$ is a three component unit vector. The first term is that of the $O(3)$-sigma model, and the second term is the baby Skyrme term with constant coefficient $\kappa^2$. The third term is a potential term containing the pion mass parameter $m$ (named in analogy with the pion mass term in the full Skyrme model), and a constant unit vector $\pmb{n}$. Greek indices run over spacetime coordinates $t$, $x$ and $y$, and later we will use Latin indices to run over the purely spatial coordinates. The symmetry of the action is $O(3)$ for $m=0$ and $O(2)$ for $m\ne 0$.

The associated static energy of this model is
\begin{align}\label{BSE}
E_{BS} = \frac{1}{2}\int\!\frac{1+r^2}{1-r^2}\Big( |\partial_x\pmb{\phi}|^2 + |\partial_y\pmb{\phi}|^2 + \frac{\kappa^2(1-r^2)^2}{4 L^2}& |\partial_x\pmb{\phi}\times\partial_y\pmb{\phi}|^2 \nonumber\\
&+ \frac{8 L^2 m^2}{(1-r^2)^2}(1-\pmb{\phi}\cdot\pmb{n})\Big)\, dx\, dy\, ,
\end{align}
which yields the equations of motion
\begin{align}\label{EOM}
\partial_i\pmb{j}_i &\equiv\partial_i \left\{\frac{1+r^2}{1-r^2}\left(\pmb{\phi}\times\partial_i\pmb{\phi} + \frac{\kappa^2(1-r^2)^2}{4 L^2}\partial_j\pmb{\phi}(\pmb{\phi}\cdot\partial_i\pmb{\phi}\times\partial_j\pmb{\phi})\right)\right\}\nonumber\\ &= m^2 \sqrt{-\det{g}}\,\pmb{n}\times\pmb{\phi}\,.
\end{align}

For finite energy we then require $\pmb{\phi} \rightarrow \pmb{n}$ as $r\rightarrow 1$. Without loss of generality we can choose $\pmb{n} = (0,0,1)$, and in the massless ($m=0$) case this choice of boundary value breaks the $O(3)$ symmetry of the model to $O(2)$. We can identify points on the boundary $r=1$ and treat the pion fields as maps $\pmb{\phi}:S^2\to S^2$, giving rise to an associated winding number and topological charge
\begin{equation}
B = -\frac{1}{4\pi}\int\!\pmb{\phi}\cdot(\partial_x\pmb{\phi}\times\partial_y\pmb{\phi})\, dx\, dy
\end{equation}
which we identify with the baryon number of the configuration. By noting the inequalities
\begin{equation}\label{Bog}
\left|\partial_x\pmb{\phi} \pm \pmb{\phi}\times\partial_y\pmb{\phi} \right|^2\ge 0\, ,\quad \frac{1+r^2}{1-r^2}\ge 1\, ,
\end{equation}
we obtain the Bogomolny bound $E_{BS}\ge 4\pi |B|$.

\section{Radial baby Skyrmions in AdS}
We begin by discussing some properties of radially symmetric solitons in our model, working in global coordinates \eqref{metricglob}. Due to the principle of symmetric criticality and the symmetries of both AdS and the action, we would expect static $B=1$ solitons in our model to posses radial symmetry and be centred at the origin $\rho=0$. Static, radially symmetric configurations with topological charge $B$ are given by the hedgehog ansatz
\begin{equation}\label{hedgehog}
\pmb{\phi} = (\sin{f(\rho)}\cos{(B(\theta-\psi))}, \sin{f(\rho)}\sin{(B(\theta-\psi))},\cos{f(\rho)})
\end{equation}
where $f(\rho)$ is some profile function satisfying $f(0)=\pi$, $f(\infty)=0$. $\psi$ is some constant internal phase which we can set to zero here due to symmetry, although internal phase differences will become important when we consider multi-solitons. Substituting into \eqref{BSE} and performing the coordinate transformation yields the static energy
\begin{equation}\label{radE}
E_{rad} = \frac{L\pi}{2}\int_{0}^{\infty}\!\sinh{\frac{2\rho}{L}}\left({f^\prime}^2 + \frac{B^2\sin^2\!{f}}{L^2\sinh^2\!{\frac{\rho}{L}}}(1+\kappa^2{f^\prime}^2) + 2m^2(1-\cos{f}) \right)\, d\rho\,.
\end{equation}
%and equation of motion
%\begin{equation}\label{feom}
%\partial_\rho\left(\sinh{\frac{2\rho}{L}}\left(L^2 + \frac{B^2 \kappa^2\sin^2\!{f}}{\sinh^2\!{\frac{\rho}{L}}}\right)f^\prime\right) =\sinh{\frac{2\rho}{L}}\sin{f}\left(\frac{B^2\cos{f}}{\sinh^2\!{\frac{\rho}{L}}}(1+\kappa^2{f^\prime}^2) + m^2 L^2\right)\, .
%\end{equation}

We can numerically find the profile functions $f(\rho)$ for different values of $B$ by minimising \eqref{radE} using a modified gradient flow method.
%Below, Figure~\ref{figradials} displays the numerical results for topological charges $1\le B \le 3$ and parameter values $\kappa=0.1$, $L=1$, where we have used fourth-order accurate approximations for derivatives on a grid with $2001$ points. As a comparison we can also perform full-field numerical minimisations on \eqref{BSE} and extract the profile function associated with radially symmetric fields. By using a hedgehog ansatz as an initial condition and performing a gradient flow mechanism with fourth-order accurate derivatives on a grid with $501\times501$ gridpoints we find radial energy minima for charges $1\le B \le 3$ that are identical to the numerical solutions to \eqref{feom}.

In other baby Skyrme models it has been found that radial solitons can be well-approximated by flat-space instantons of the $O(3)$-sigma model. This approximation has enabled an investigation of how the soliton sizes $\mu$ scale with the baby Skyrme parameter $\kappa$. Unfortunately $O(3)$-sigma instantons in AdS do not have a convenient closed form expression, due to the presence of the non-constant time component of the metric, preventing us from analytically exploring this relationship. We can nevertheless perform a numerical investigation; since the profile function interpolates between $\pi$ and $0$ we can define the size of the soliton as $\mu : f(\mu)=\pi/2$.

Using this definition it is straightforward to numerically find how $\mu$ scales with $\kappa$ and $L$ for different values of $B$.
%as can be seen in Figure~\ref{figsizes}
In the massless case we find the leading-order dependence is $\mu \sim \sqrt{\kappa L}$ for small $\kappa/L$; looking at \eqref{radE} we see that nonlinear effects will dominate when $\kappa/L$ is large. Comparing this to the scaling of baby Skyrmions with mass parameter $m>0$ in flat space, $\mu\sim\sqrt{\kappa/m}$, we can see that the curvature of AdS space can be interpreted as adding an effective pion mass.

%\begin{figure}[p]
%	\centering
%	\includegraphics[width=0.5\textwidth]{radial3.eps}
%	\caption{Numerically computed radial AdS baby Skyrmion profile functions for topological charges $1\le B\le 3$, with $\kappa=0.1$ and $L=1$.}
%	\label{figradials}
%\end{figure}
%%
%\begin{figure}[p]
%	\centering
%	\begin{subfigure}[b]{0.4\textwidth}
%		\includegraphics[width=\textwidth]{fixedL.eps}
%        \label{fixedL}
%	\end{subfigure}
%	~
%	\begin{subfigure}[b]{0.4\textwidth}
%		\includegraphics[width=\textwidth]{fixedK.eps}
%        \label{fixedK}
%	\end{subfigure}
%	\hfill
%	\caption{Log-log plots demonstrating how the size $\mu$ varies with the baby Skyrme parameter $\kappa$ (left, with fixed $L=1$) and the AdS radius $L$ (right, with fixed $\kappa=0.1$) for radial AdS baby Skyrmions with $1\le B\le 3$. The dashed lines have gradient $0.5$, showing that the sizes scale approximately as $\mu\sim\sqrt{\kappa L}$ for small $\kappa/L$.}
%	\label{figsizes}
%\end{figure}

Finally, we can calculate the leading-order decay of the soliton tails near the boundary of our space. Assuming a radial ansatz, and using the fact that $f(\rho)\to 0$ as $\rho\to\infty$, we can linearise the equations of motion to obtain
\begin{equation}
L^2\sinh{\frac{2\rho}{L}}f^{\prime\prime} + 2L\cosh{\frac{2\rho}{L}}f^\prime - \sinh{\frac{2\rho}{L}}\left(\frac{B^2}{\sinh^2\!{\frac{\rho}{L}}}+m^2 L^2\right)f = 0\,.
\end{equation}
In the limit $\rho\to\infty$ we can obtain the asymptotic tail decay as
\begin{equation}
f(\rho)\sim e^{-\left(1+\sqrt{1+m^2 L^2}\right)\rho/L}\,,
\end{equation}
independently of $\kappa$ or $B$. This relation also holds in the special case $m=0$.
%This relation can also be verified numerically as shown in Figure~\ref{figdecay} where the large $r$ behaviour of the profile function is explored for different values of $\kappa$ and $B$.
It is interesting to note that, unlike baby Skyrme models in flat space, the addition of a mass term is not required for the soliton tail to decay exponentially. This is in contrast with baby Skyrmions in flat space which have large-radius asymptotic tail decays
\begin{equation}
f(r)\sim\begin{cases}
r^{-B}, & \text{if $m=0$}\\
\frac{1}{\sqrt{r}}e^{-mr}, & \text{if $m\ne 0$}\,.
\end{cases}
\end{equation}
In fact, these flat-space tail decays can be obtained from the linearised asymptotic equations of motion by carefully taking the limit $L\to\infty$, as expected.

%\begin{figure}[t]
%	\centering
%	\includegraphics[width=0.5\textwidth]{decay.eps}
%	\caption{A log-log plot of the profile function $f(r)$ against $(1-r)^2$ for radial AdS baby Skyrmions with $1\le B\le 3$. For large $r$ we clearly see the profile functions decay as $f(r)\sim (1-r)^2$.}
%	\label{figdecay}
%\end{figure}

\begin{figure}[p]
	\centering
	%    \missingfigure[figwidth=8cm]{Graph waiting for numerical simulations to finish}
	\begin{subfigure}[b]{0.23\textwidth}
		\includegraphics[width=\textwidth]{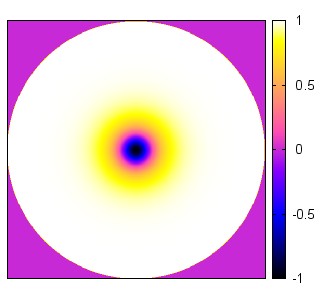}\caption{$B=1$}
	\end{subfigure}
	~
	\begin{subfigure}[b]{0.23\textwidth}
		\includegraphics[width=\textwidth]{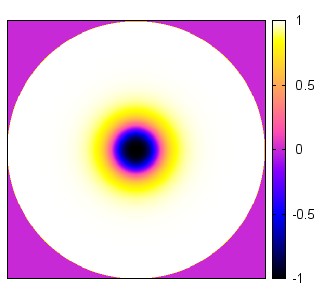}\caption{$B=2$}
	\end{subfigure}
	~
	\begin{subfigure}[b]{0.23\textwidth}
		\includegraphics[width=\textwidth]{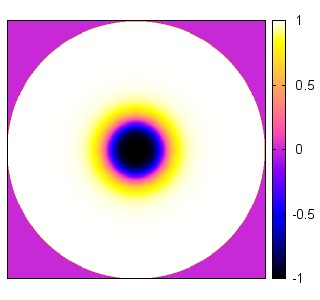}\caption{$B=3$}
	\end{subfigure}
	~
	\begin{subfigure}[b]{0.23\textwidth}
		\includegraphics[width=\textwidth]{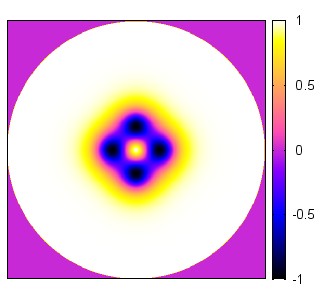}\caption{$B=4$}
	\end{subfigure}
	\hfill\\
	\begin{subfigure}[b]{0.23\textwidth}
		\includegraphics[width=\textwidth]{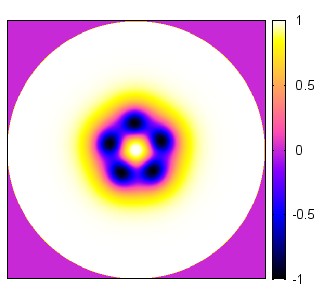}\caption{$B=5$}
	\end{subfigure}
	~
	\begin{subfigure}[b]{0.23\textwidth}
		\includegraphics[width=\textwidth]{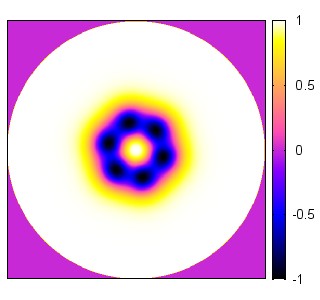}\caption{$B=6$}
	\end{subfigure}
	~
	\begin{subfigure}[b]{0.23\textwidth}
		\includegraphics[width=\textwidth]{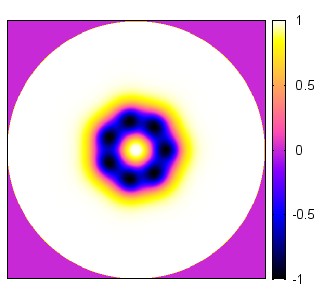}\caption{$B=7$}
	\end{subfigure}
	~
	\begin{subfigure}[b]{0.23\textwidth}
		\includegraphics[width=\textwidth]{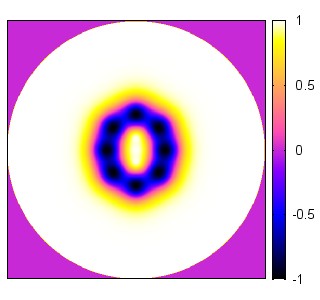}\caption{$B=8$}
	\end{subfigure}
	\hfill\\
	\begin{subfigure}[b]{0.23\textwidth}
		\includegraphics[width=\textwidth]{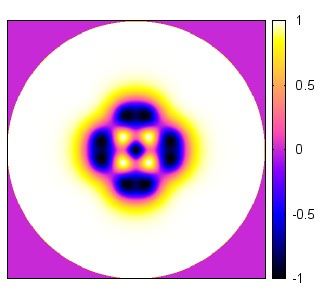}\caption{$B=9$}
	\end{subfigure}
	~
	\begin{subfigure}[b]{0.23\textwidth}
		\includegraphics[width=\textwidth]{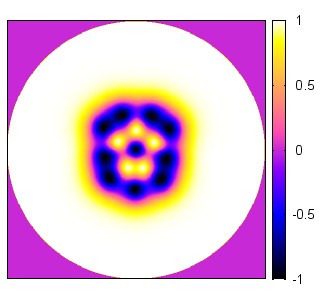}\caption{$B=10$}
	\end{subfigure}
	~
	\begin{subfigure}[b]{0.23\textwidth}
		\includegraphics[width=\textwidth]{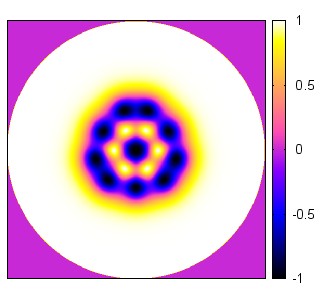}\caption{$B=11$}
	\end{subfigure}
	~
	\begin{subfigure}[b]{0.23\textwidth}
		\includegraphics[width=\textwidth]{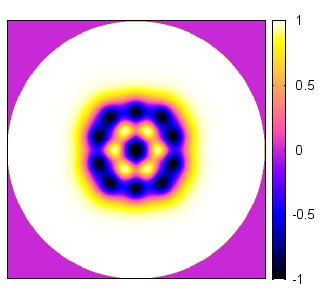}\caption{$B=12$}
	\end{subfigure}
	\hfill\\
	\begin{subfigure}[b]{0.23\textwidth}
		\includegraphics[width=\textwidth]{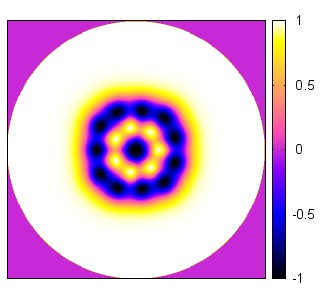}\caption{$B=13$}
	\end{subfigure}
	~
	\begin{subfigure}[b]{0.23\textwidth}
		\includegraphics[width=\textwidth]{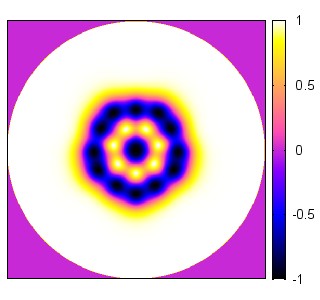}\caption{$B=14$}
	\end{subfigure}
	~
	\begin{subfigure}[b]{0.23\textwidth}
		\includegraphics[width=\textwidth]{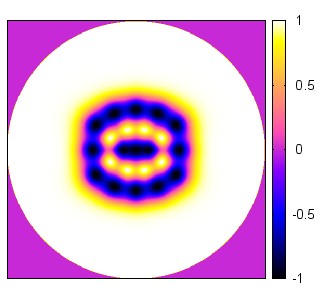}\caption{$B=15$}
	\end{subfigure}
	~
	\begin{subfigure}[b]{0.23\textwidth}
		\includegraphics[width=\textwidth]{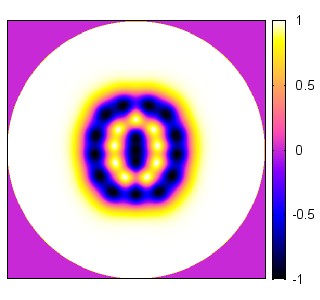}\caption{$B=16$}
	\end{subfigure}
	\hfill\\
	\begin{subfigure}[b]{0.23\textwidth}
		\includegraphics[width=\textwidth]{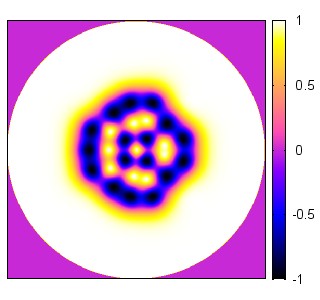}\caption{$B=17$}
	\end{subfigure}
	~
	\begin{subfigure}[b]{0.23\textwidth}
		\includegraphics[width=\textwidth]{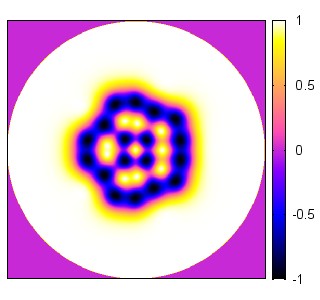}\caption{$B=18$}
	\end{subfigure}
	~
	\begin{subfigure}[b]{0.23\textwidth}
		\includegraphics[width=\textwidth]{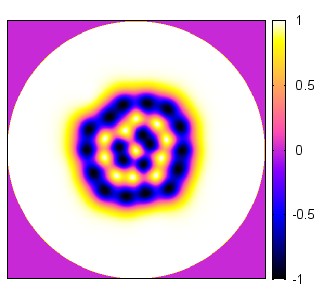}\caption{$B=19$}
	\end{subfigure}
	~
	\begin{subfigure}[b]{0.23\textwidth}
		\includegraphics[width=\textwidth]{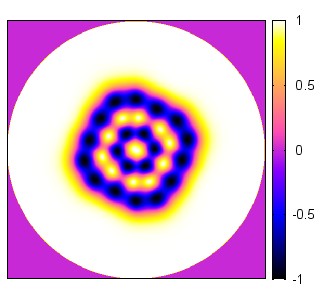}\caption{$B=20$}
	\end{subfigure}
	\hfill\\
	\caption{Plots of the numerically calculated AdS baby Skyrmions for topological charges $1\le B\le 20$ for $\kappa=0.1$, $L=1$ and $m=0$. We plot $\phi_3$ in sausage coordinates as a visual representation of the solitons.}
	\label{adsbabies}
\end{figure}

\section{Multi-solitons in AdS}
As we shall see, for topological charges $B>3$, radial baby Skyrmions in AdS no longer provide global minima for the energy functional \eqref{BSE}. Analytic investigation of higher charge solitons is difficult, but we are able to perform full numerical minimisations to seek out local and global energy minima. The numerical results in this section were obtained by performing a modified gradient flow method on the energy \eqref{BSE} with parameter values $\kappa=0.1$, $L=1$, $m=0$ on a grid with $501\times501$ gridpoints in sausage coordinates. Derivatives were calculated used fourth-order finite difference approximations. In addition, the numerical results were verified by applying a fourth-order Runge-Kutta method to the full dynamical equations of motion.

In order to find local energy minima we investigated a range of different initial conditions for our minimisation algorithms. Initial conditions were primarily generated using the product ansatz: if we project a field $\pmb{\phi}$ onto the Riemann sphere by
\begin{equation}
W = \frac{\phi_1 + i\phi_2}{1-\phi_3}\,,
\end{equation}
then we can generate a field configuration composed from two fields $W_1$, $W_2$ by writing
\begin{equation}
W = \frac{W_1 W_2}{W_1 + W_2}\,.
\end{equation}
Positions of solitons can be identified with points where $\pmb{\phi}=(0,0,-1)$ or equivalently $W=0$, and $W=\infty$ whenever $\pmb{\phi}=(0,0,1)$, the boundary value. Our composed field has zeroes at the zeroes of $W_1$ and $W_2$, so we see that the product ansatz gives us a field with the correct topological properties as the superposition of two individual fields. Combining this with the hyperbolic translations \eqref{ADStrans} allows us to generate initial conditions that resemble collections of radial baby Skyrmions placed in different positions on our grid. We also used perturbed radial fields as initial conditions.

Figure~\ref{adsbabies} shows colour contour plots of $\phi_3$ for numerically found global energy minima for topological charges $1\le B\le 20$, with energies given in Table~\ref{babytable}. For the pion mass term used in this paper, radial solutions are preferred up to charge $B=3$, in contrast to baby Skyrmions in flat space which only have radial energy minima for $B\le 2$.

\begin{table}[b]
	\centering
	\begin{tabular}{ | c c c || c c c | }
		\hline
		charge $B$ & $E/4\pi B$ & form & charge $B$ & $E/4\pi B$ & form\\
		\hline                       
		1 & 1.2548 & \{1\} &	11 & 1.5368 & \{2, 9\}  \\
		2 & 1.2312 & \{2\} &	12 & 1.5554 & \{2, 10\}  \\
		3 & 1.2878 & \{3\} &	13 & 1.5788 & \{2, 11\}  \\
		4 & 1.3384 & \{4\} &	14 & 1.6017 & \{2, 12\} \\
		5 & 1.3725 & \{5\} &	15 & 1.6250 & \{3, 12\} \\
		6 & 1.3886 & \{6\} &	16 & 1.6481 & \{3, 13\} \\
		7 & 1.4263 & \{7\} &	17 & 1.6714 & \{4, 13\} \\
		8 & 1.4541 & \{8\} &	18 & 1.6914 & \{4, 14\} \\
		9 & 1.4888 & \{1, 8\} &	19 & 1.7107 & \{5, 14\} \\
		10 & 1.5157 & \{1, 9\} &	20 & 1.7276 & \{6, 14\} \\
		
		\hline
	\end{tabular}
	\caption{Energies per charge and forms for AdS baby Skyrmions with topological charge $1\le B\le 20$.}
	\label{babytable}
\end{table}

For $4\le B\le 7$ the baby Skyrmions form regular polygons, with soliton positions at the vertices. The relative phase differences between neighbouring soliton positions is $\pi$ or $\pi\pm\pi/B$ for even or odd charges respectively.

At charge $B=8$ the ring structure deforms due to the centralising force induced by the AdS metric. For $B\ge9$ the solutions form multi-layered concentric rings, and the central layers for $9\le B\le 16$ resemble (normally slightly deformed) radial solutions. These deformations appear to match the symmetries of the outer rings. We denote multi-layered ring structures as $\{n_1, n_2, n_3, \dots\}$, where $n_i$ denotes the topological charge in the $i$th ring, counting out from the origin.

The transition from a single ring to a multi-ring structure with increasing baryon number is reminiscient of the popcorn transition observed in low-dimensional analogues of the Sakai-Sugimoto \cite{Bolognesi:2013jba, Elliot-Ripley:2015cma} model of holographic QCD, in which finite density chains of solitons pop out into a holographic direction at some critical baryon density. In light of this, it would be interesting to see if the AdS baby Skyrme model studied here possesses further popcorn-like transitions beyond $B=20$.

This is a potentially difficult task to accomplish numerically. As higher charge solutions are investigated the number of local energy minima increases drastically. This requires a very large number of initial conditions to be tested in order to gain confidence that a global energy minimum has indeed been found.

In the following sections we will formulate an approximation to our system in which the solitons are modelled by point particles in a gravitational potential with an inter-soliton interaction. We will use this model to predict the baryon numbers at which further popcorn transitions occur, and use the predictions as a guide to finding further global minima in the full model.

\section{AdS baby Skyrmions as point particles}

The numerical results from the previous section are reminiscent of the results of circle packings within a circle \cite{graham1998dense}: finding the minimal area circle within which you can pack $B$ congruent circles. Solutions tend to be arranged as rings of points separated by at least the diameter of the circles ($\sim 2\mu$ for our baby Skyrmions). However, using this as an approximation to the AdS baby Skyrme model presents some problems. Firstly, the popcorn transitions occur too early, the first two at $B=7$ and $B=19$. This is likely due to the malleable and overlapping nature of the baby Skyrmions. This problem still occurs even when the circle packings are formulated in hyperbolic space. However it does suggest that a point-particle approximation could be able to qualitatively predict the form of solutions, if a better representation of the soliton interactions can be found.

In order to derive a better point-particle approximation it will be necessary to obtain numerical approximations to the effective gravitational potential and inter-soliton interaction. We will make use of the hyperbolic translations \eqref{ADStrans}, and assume that the energies of radial fields translated in this way can approximate the energies of constituent parts of multi-soliton configurations located at different points on our grid. It should be clarified that such fields are not solutions to the equations of motion since the hyperbolic translations are not isometries of AdS.

\subsection{The gravitational potential}
We begin by deriving an approximate gravitational potential from the metric \eqref{metric}. The geodesic equations associated with this metric are given by
\begin{equation}\label{geodesic}
\begin{aligned}
t^{\prime\prime}&=-\frac{8}{1-r^4}(xx^\prime+yy^\prime)t^\prime\,,\\[5pt]
x^{\prime\prime}&=-\frac{x(1+r^2)}{L^2(1-r^2)}(t^\prime)^2 + \frac{2x}{1-r^2}(y^\prime)^2 - \frac{4x}{1-r^2}x^\prime y^\prime\,,\\[5pt]
y^{\prime\prime}&=-\frac{y(1+r^2)}{L^2(1-r^2)}(t^\prime)^2 + \frac{2y}{1-r^2}(x^\prime)^2 - \frac{4y}{1-r^2}x^\prime y^\prime\,,
\end{aligned}
\end{equation}
where primes denote differentiation with respect to proper time. In the non-relativistic limit $x^\prime$, $y^\prime\ll t^\prime$ we can write
\begin{equation}
\ddot{x} = \frac{x^{\prime\prime}}{(t^\prime)^2} - \frac{x^\prime t^{\prime\prime}}{(t^\prime)^3}\approx \frac{x^{\prime\prime}}{(t^\prime)^2}\approx -\frac{x(1+r^2)}{L^2(1-r^2)}\equiv -\frac{x}{r}\partial_r \Phi\,,
\end{equation}
where we have implicitly defined the gravitational potential $\Phi(r)$. Integrating along with the condition $\Phi(0)=0$ gives
\begin{equation}
\Phi(r) = \int_{0}^{r}\!\frac{R(1+R^2)}{L^2(1-R^2)}\,dR = -\frac{1}{2L^2}(r^2+2\log{(1-r^2)})\,.
\end{equation}

To fit this potential to the AdS baby Skyrme model we are required to multiply it by some constant factor $\alpha$. We obtain a numerical approximation for the gravitational potential by evaluating the energies of translated $B=1$ radial fields and subtracting off the energy of the true $B=1$ solution. We then fit $\alpha$ to this data by performing a least-squares fit. We fit only within the radius $r=0.6$ because full numerical local minima, even for high charges, do not lie much beyond this radius, and the hyperbolic translated $B=1$ fields become less accurate as approximations near the edge of the disc. In addition, we investigated different radii to fit our data to, and found that choosing values larger than $0.6$ resulted in point-particle approximations that did not successfully estimate the full numerical results.

\begin{figure}[t]
	\centering
	\includegraphics[width=0.5\textwidth]{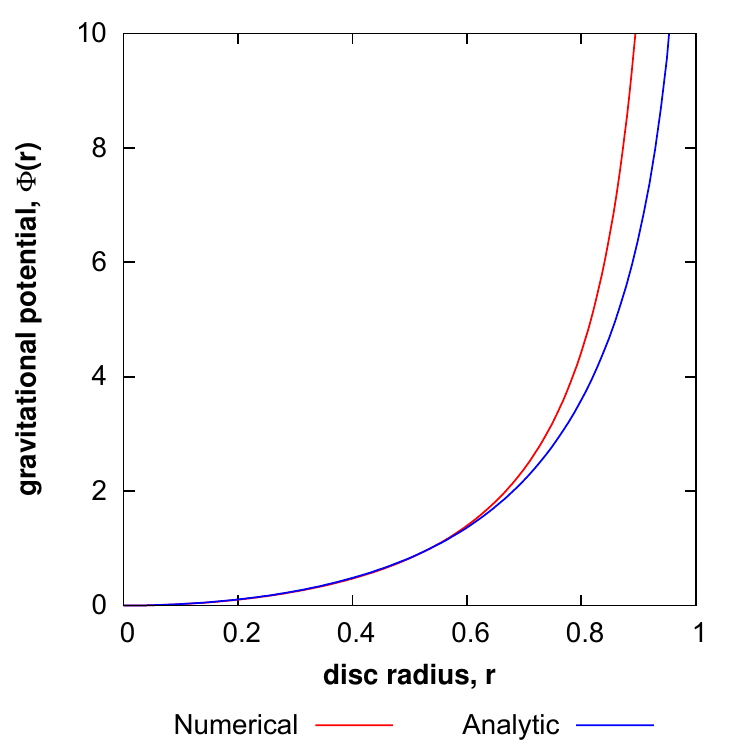}
	\caption{Numerical and analytical approximations for the gravitational potential induced by the AdS metric. The analytic approximation is $\Phi(r) = -\alpha(r^2+2\log{(1-r^2)})/2L^2$ with $\alpha = 64.3$ and $L=1$. Energies are given in units of $4\pi$.}
	\label{figgravity}
\end{figure}

Figure~\ref{figgravity} shows the analytic potential for $\alpha = 64.3$ compared to the numerical approximations with $\kappa=0.1$, $L=1$. The curves are in close agreement for radii in the range $r\in[0,0.6]$, but diverge as $r$ increases further, as expected.

\subsection{The inter-soliton interaction}
We can obtain a numerical approximation for the inter-soliton interaction in a similar way: we can numerically calculate the energy of a product ansatz of two translated $B=1$ solitons, subtract off the potential energies associated with each component soliton (according to the numerical approximation above) and the energy of two $B=1$ solutions, and plot the resulting energy as a function of the geodesic separation of the positions of the solitons.

\begin{figure}[p]
	\centering
	\includegraphics[width=0.3\textwidth]{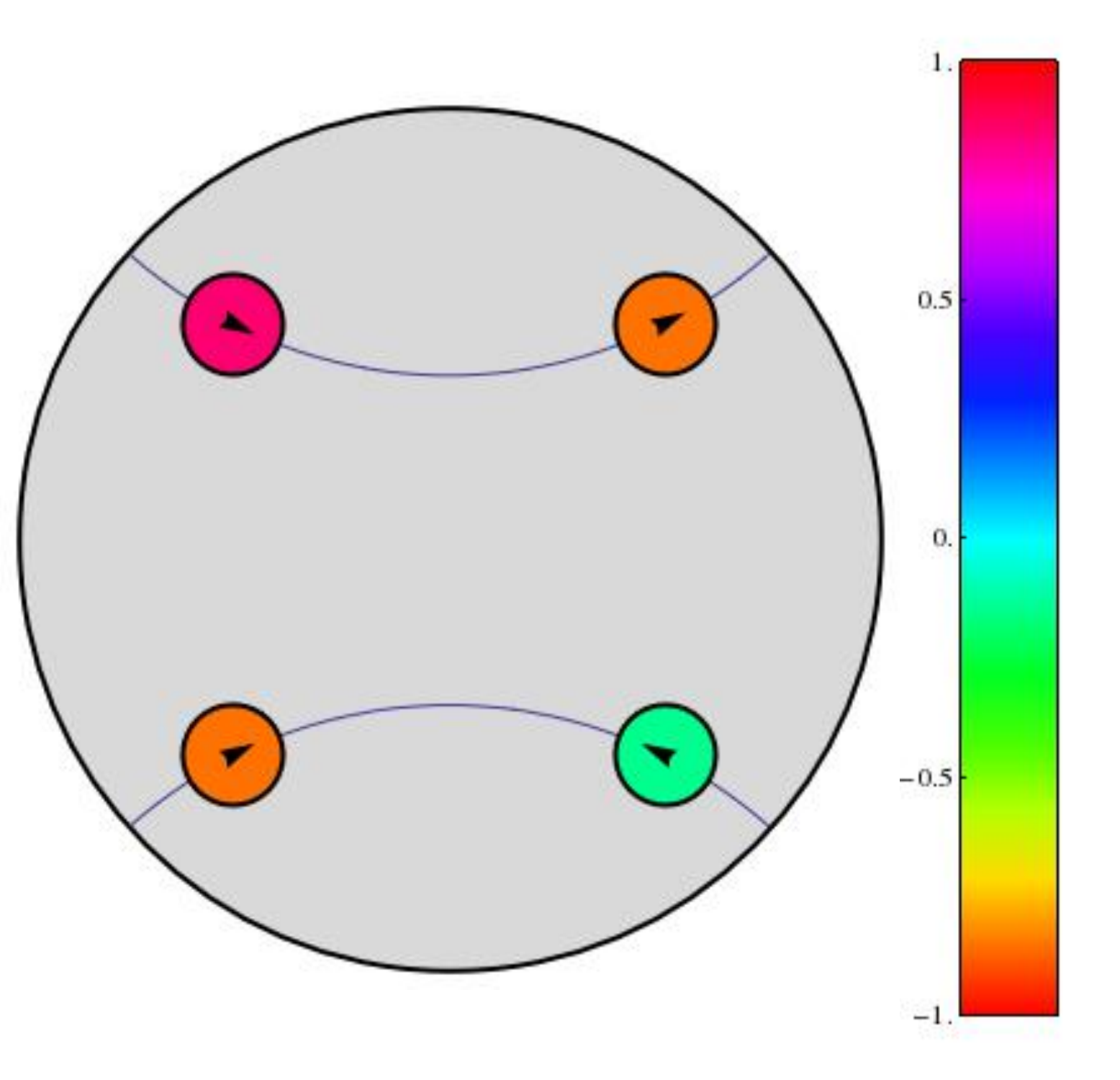}
	\caption{Diagram illustrating the geodesics connecting pairs of particles on the disc, with the top pair being in phase (the maximally repulsive channel) and the bottom pair being out of phase (the maximally attractive channel). Particles are coloured according to their internal phase, using the bar on the right, with units given in multiples of $\pi$.}
	\label{phases}
\end{figure}

\begin{figure}[p]
	\centering
	\includegraphics[width=0.5\textwidth]{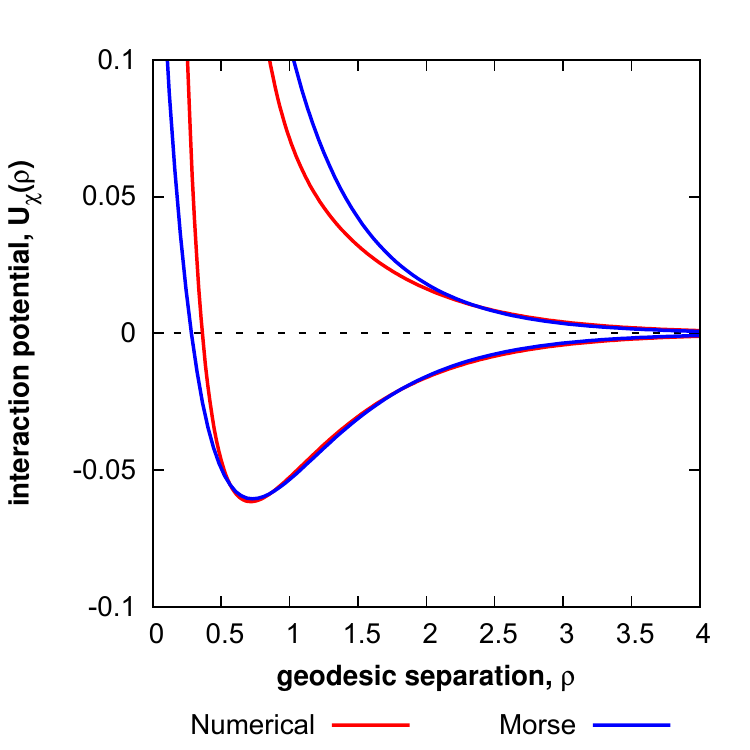}
	\caption{Numerical and analytic approximations for the inter-soliton interaction. The upper curves are the results for baby Skyrmions in phase, and the lower curves are for baby Skyrmions out of phase. The blue curves are given by the Morse potential \eqref{morse} with $D=0.76$, $\rho_e = 0.73$ and $a=1.13$. Energies are given in units of $4\pi$.}
	\label{figcoulomb}
\end{figure}

These static approximations are shown as red curves in Figure~\ref{figcoulomb}, where the upper curve represents the inter-soliton energy of a pair of solitons in phase, and the lower curve represents solitons out of phase. Relative phase differences are calculated with respect to the geodesics that the particles lie on (see Fig~\ref{phases}), so that a pair of soliton field with internal phases $\psi_a$ and $\psi_b$ have a relative phase difference $\chi = \chi(\psi_a,\psi_b)$. In keeping with results found in other spacetimes, we find that pairs of solitons at large separations are in the maximally repulsive channel when they are in phase ($\chi = 0$), and in the maximallFor higher topological chargesy attractive channel when they are out of phase ($\chi = \pi$).

With this numerical data as a guide we can fit the out-of-phase interaction using a Morse potential of the form
\begin{equation}
U_{\pi}(\rho) = D\left(e^{2a(1-\rho/\rho_e)}-2e^{a(1-\rho/\rho_e)}\right)\,,
\end{equation}
where $\rho$ is the geodesic separation between the solitons, $D$ is the depth of the potential at its minimum, $\rho_e$ is the separation at which the potential is minimised, and $a$ is a parameter controlling the width of the potential. These parameters can be fit by performing a least-squares fit with the numerical data. Since the product ansatz is only valid for well-separated solitons, we fit the data in the region where the separation between the solitons is greater than twice the size of a single soliton. This yields parameter values $D=0.76$, $\rho_e=0.73$ and $a=1.13$.

%However, when applied with this parameter we found our approximation did not reproduce the same transition to a 2-ring structure at charge $B=9$. After calibrating our model to generate this qualitative transition we settled on the parameter value $a=1.3$, displayed in the lower blue curve in Figure~\ref{figcoulomb}.

In order to introduce a dependence of the potential on the relative phase difference of the two solitons, $\chi$, we assume that the solitons are at their most attractive when they are out of phase, and their most repulsive when they are in phase i.e.
\begin{equation}\label{morse}
U_{\chi}(\rho) = D\left(e^{2a(1-\rho/\rho_e)}+2\cos{(\chi)}e^{a(1-\rho/\rho_e)}\right)\,.
\end{equation}
Plots of $U_0$ and $U_\pi$ are given by the upper and lower blue curves in Figure~\ref{figcoulomb} respectively.

\section{Baby Skyrmion rings and shells}
We now seek energy minima of the point-particle approximation above for various values of $B$. For a configuration of $B$ solitons with disc coordinates $\pmb{x}_a$ and internal phases $\psi_a$ (where $1\le a\le B$) we seek to minimise the energy
\begin{equation}\label{ptptclenergy}
E_B = \sum_{a=1}^{B}\left(\Phi(r_a) + \sum_{b>a}^{}U_{\chi(\psi_a,\psi_b)}(d(\pmb{x}_a,\pmb{x}_b))\right)\,,
\end{equation}
where $r_a\equiv|\pmb{x}_a|$, $\Phi(r)$ and $U_\chi(\rho)$ are the potentials given above and $d(\pmb{x}_a,\pmb{x}_b)$ is the geodesic distance \eqref{sausagedist} between points $\pmb{x}_a$ and $\pmb{x}_b$.

\begin{figure}[p]
	\centering
	%    \missingfigure[figwidth=8cm]{Graph waiting for numerical simulations to finish}
	\begin{subfigure}[b]{0.23\textwidth}
		\includegraphics[width=\textwidth]{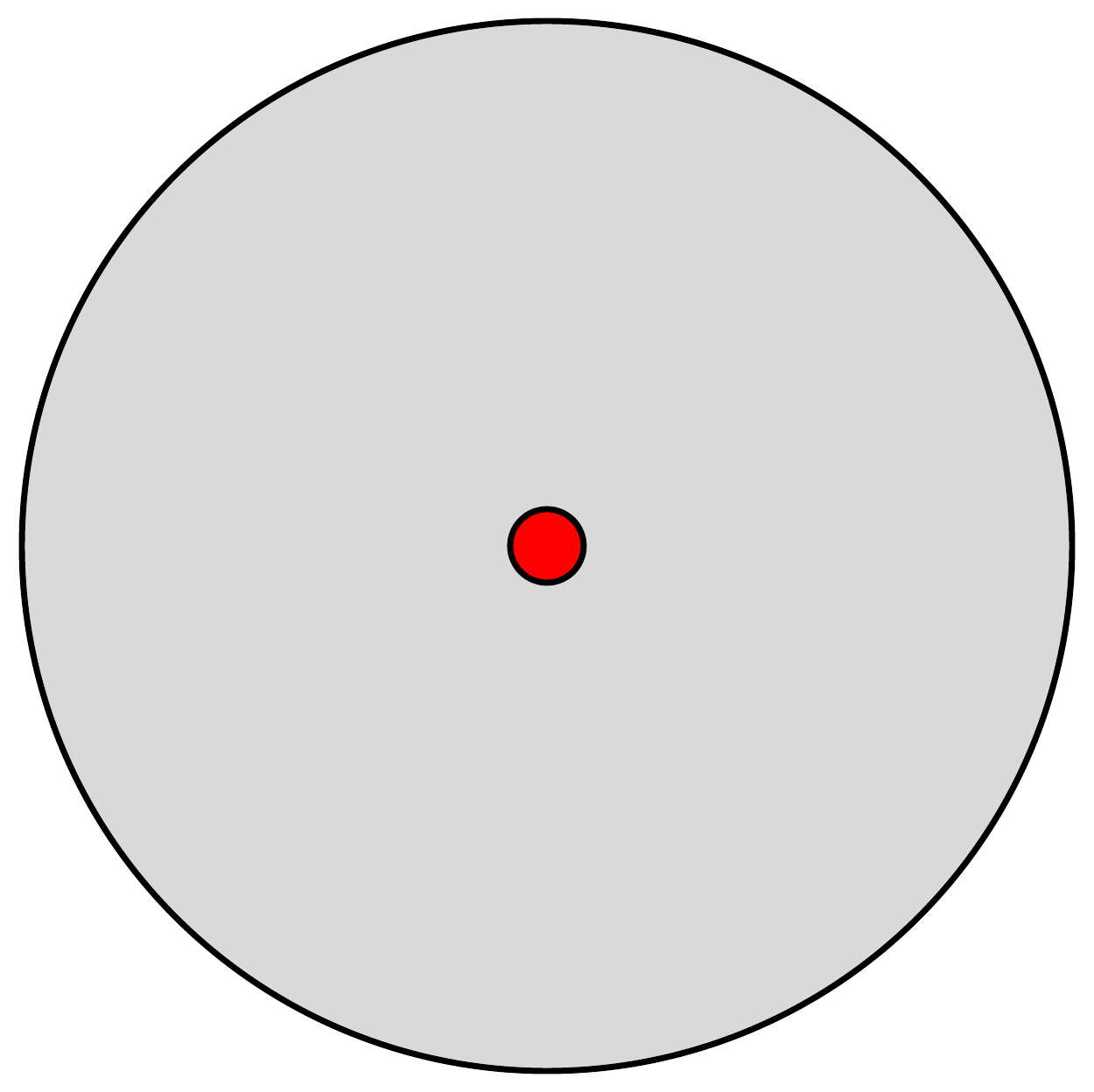}\caption{$B=1$}
	\end{subfigure}
	~
	\begin{subfigure}[b]{0.23\textwidth}
		\includegraphics[width=\textwidth]{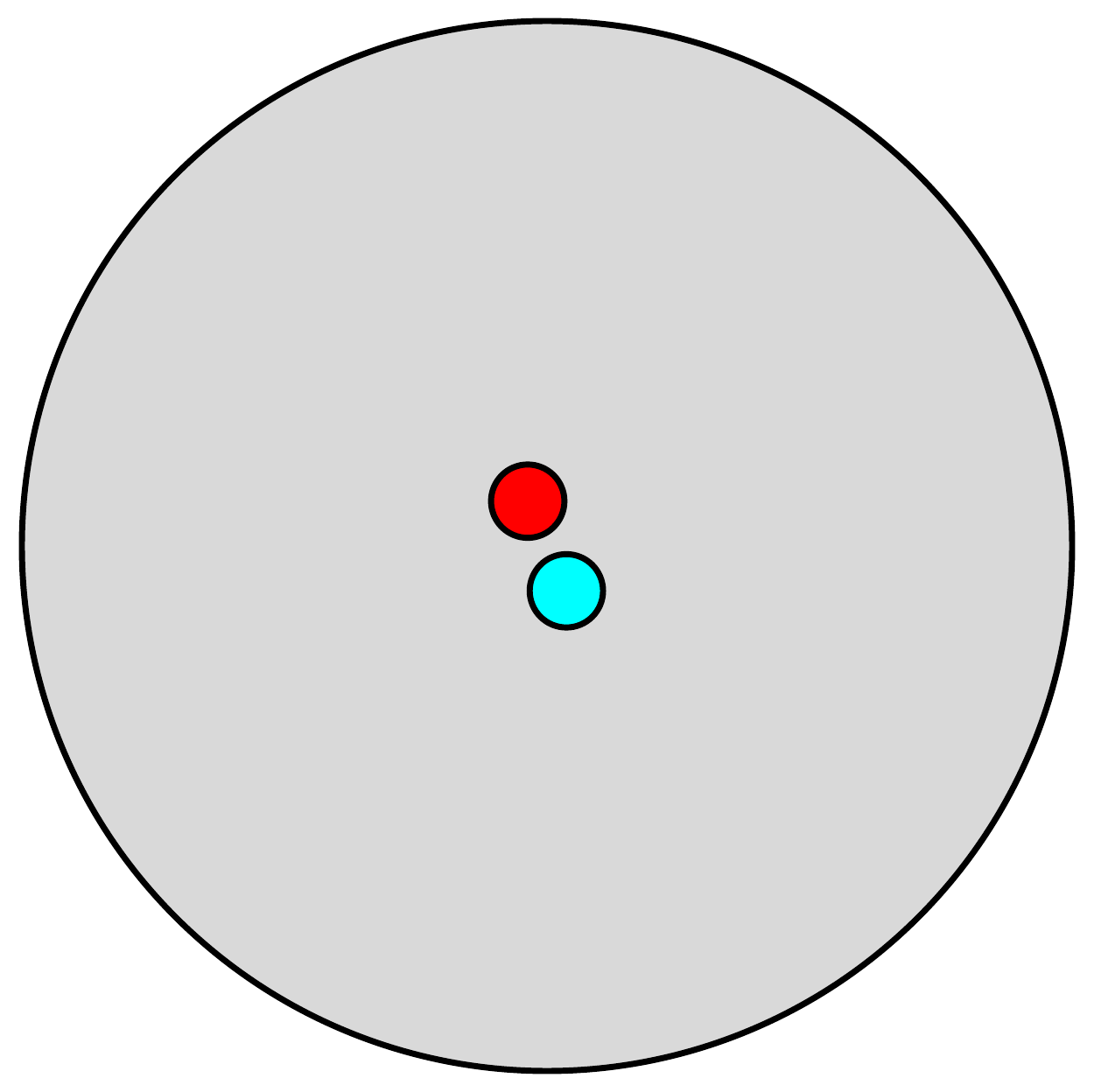}\caption{$B=2$}
	\end{subfigure}
	~
	\begin{subfigure}[b]{0.23\textwidth}
		\includegraphics[width=\textwidth]{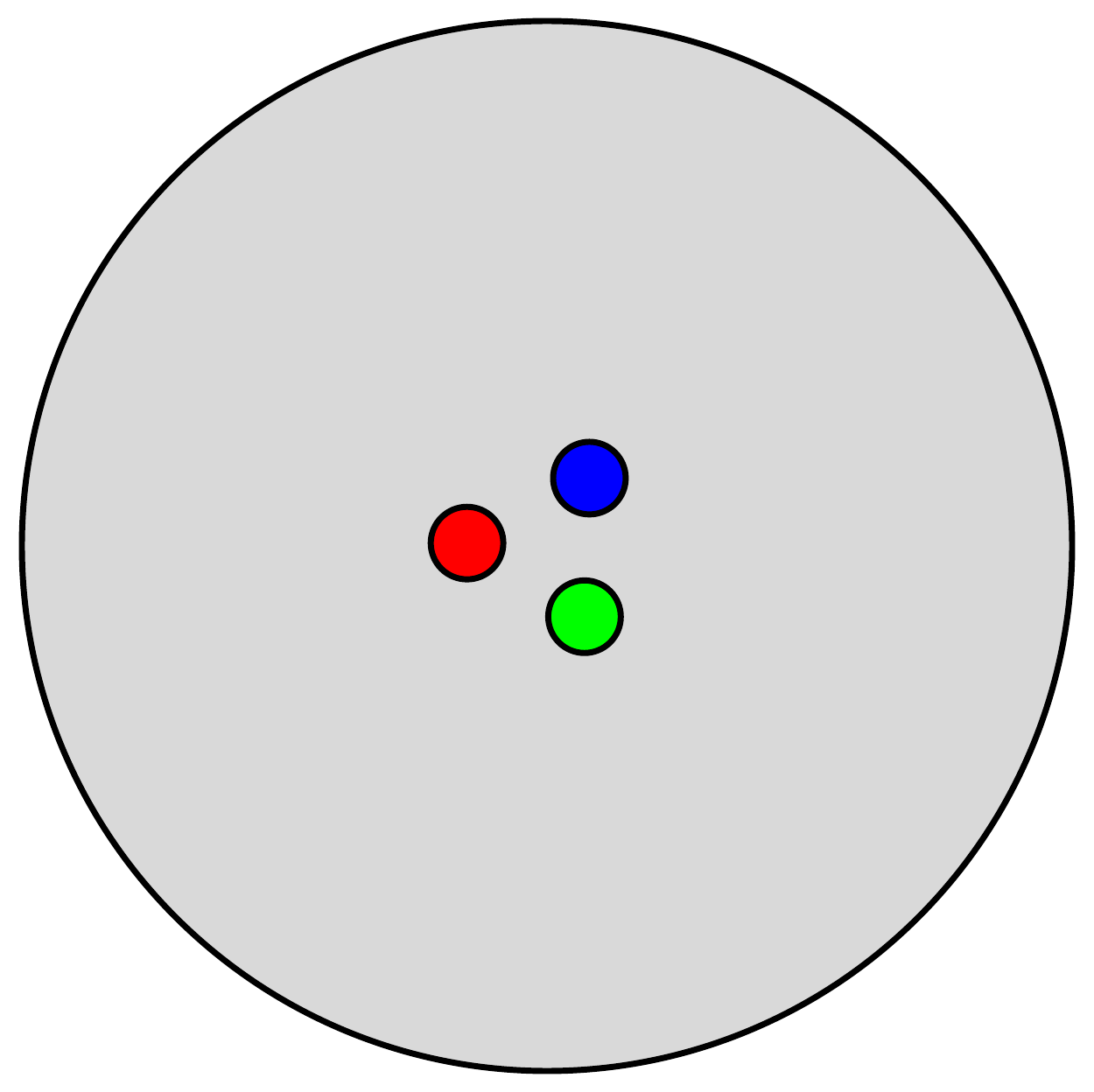}\caption{$B=3$}
	\end{subfigure}
	~
	\begin{subfigure}[b]{0.23\textwidth}
		\includegraphics[width=\textwidth]{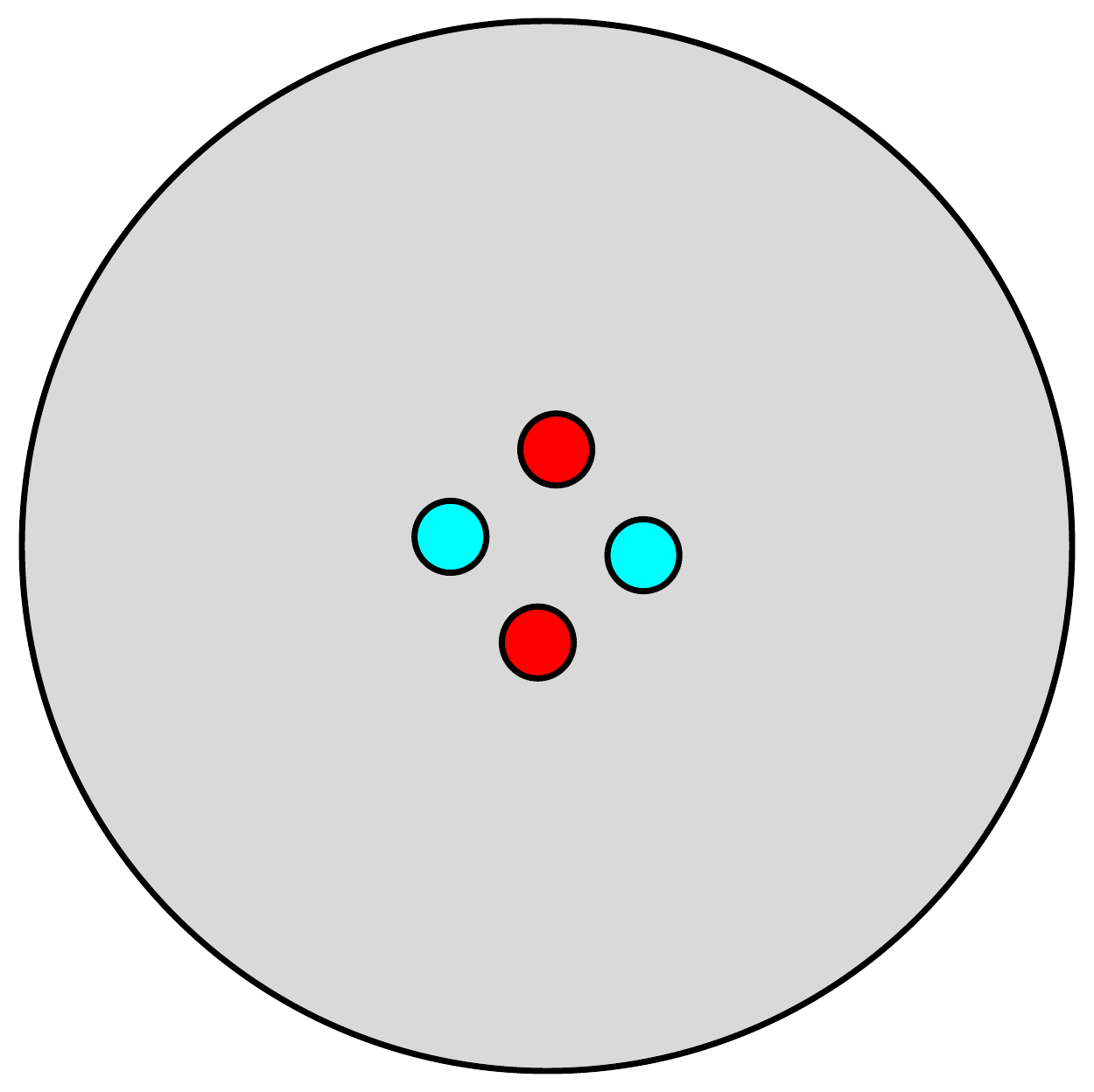}\caption{$B=4$}
	\end{subfigure}
	\hfill\\
	\begin{subfigure}[b]{0.23\textwidth}
		\includegraphics[width=\textwidth]{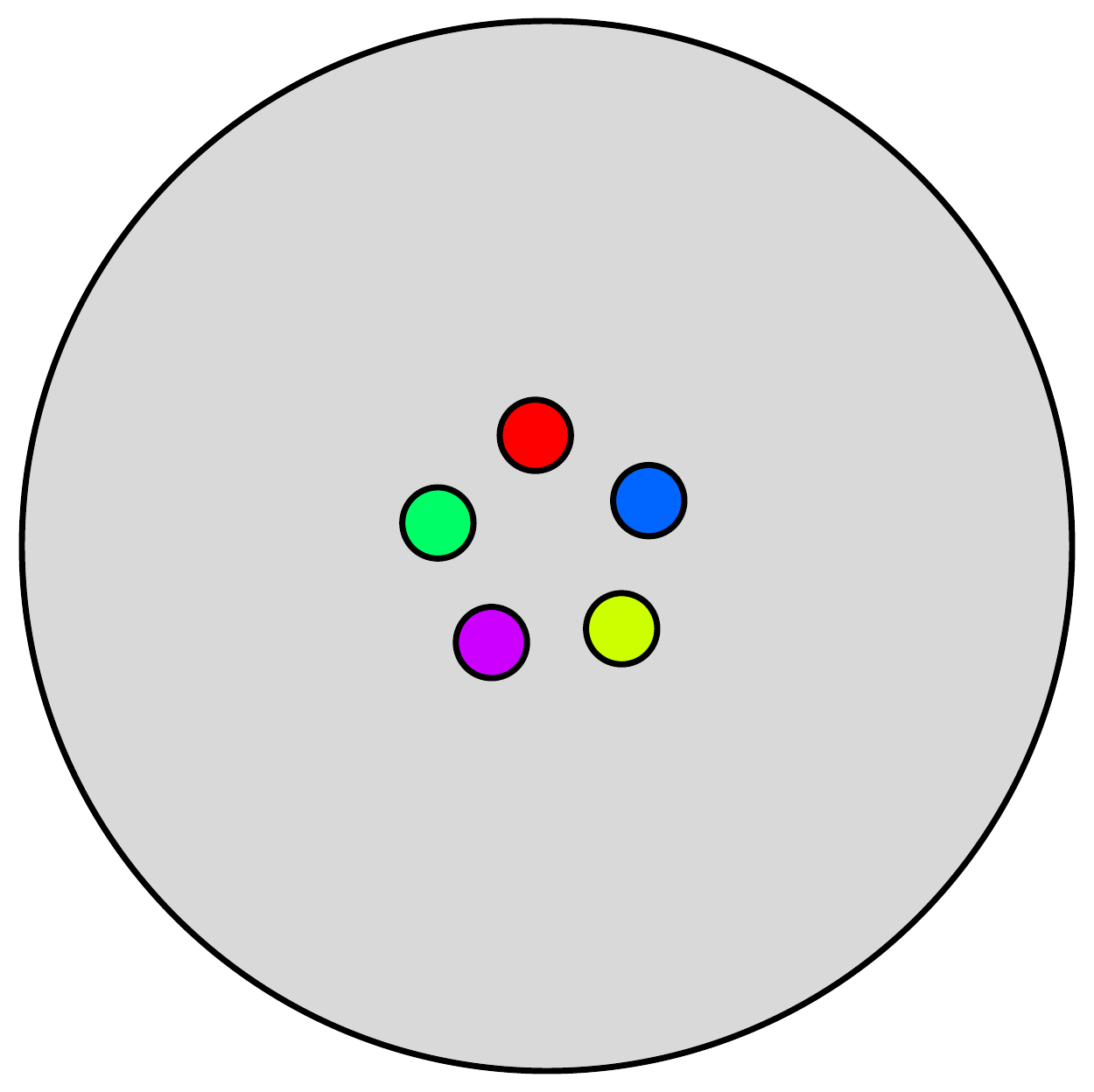}\caption{$B=5$}
	\end{subfigure}
	~
	\begin{subfigure}[b]{0.23\textwidth}
		\includegraphics[width=\textwidth]{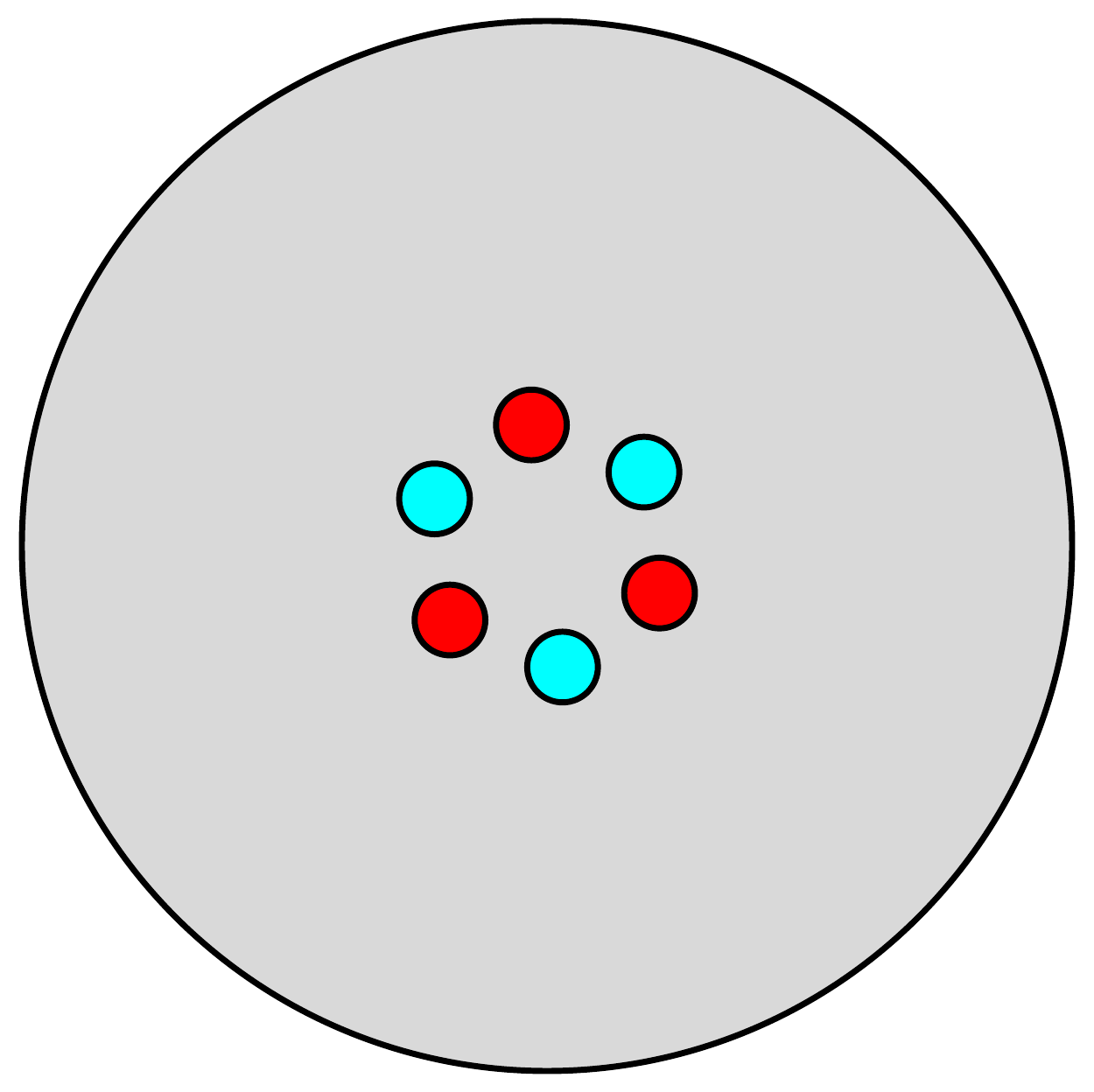}\caption{$B=6$}
	\end{subfigure}
	~
	\begin{subfigure}[b]{0.23\textwidth}
		\includegraphics[width=\textwidth]{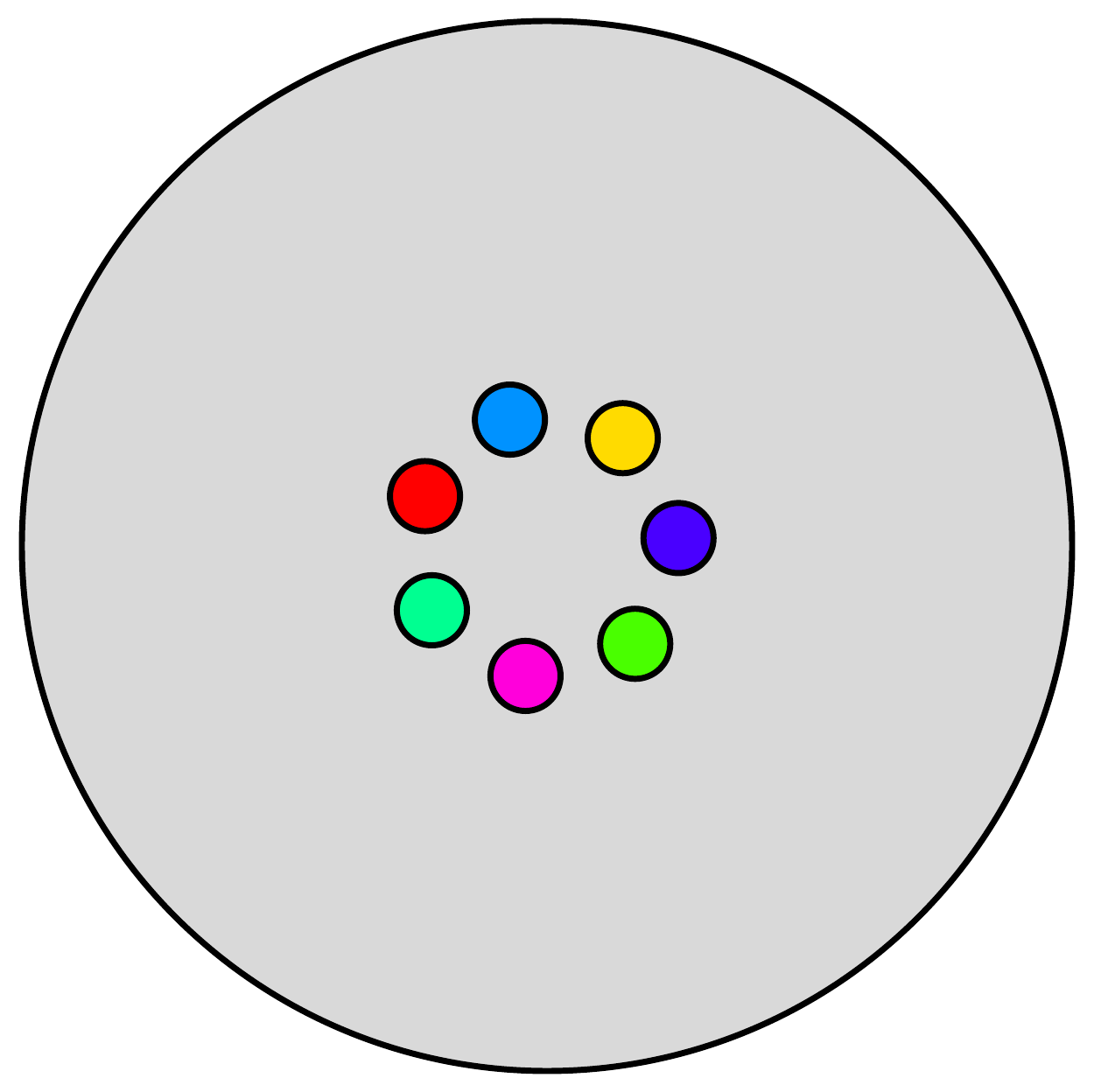}\caption{$B=7$}
	\end{subfigure}
	~
	\begin{subfigure}[b]{0.23\textwidth}
		\includegraphics[width=\textwidth]{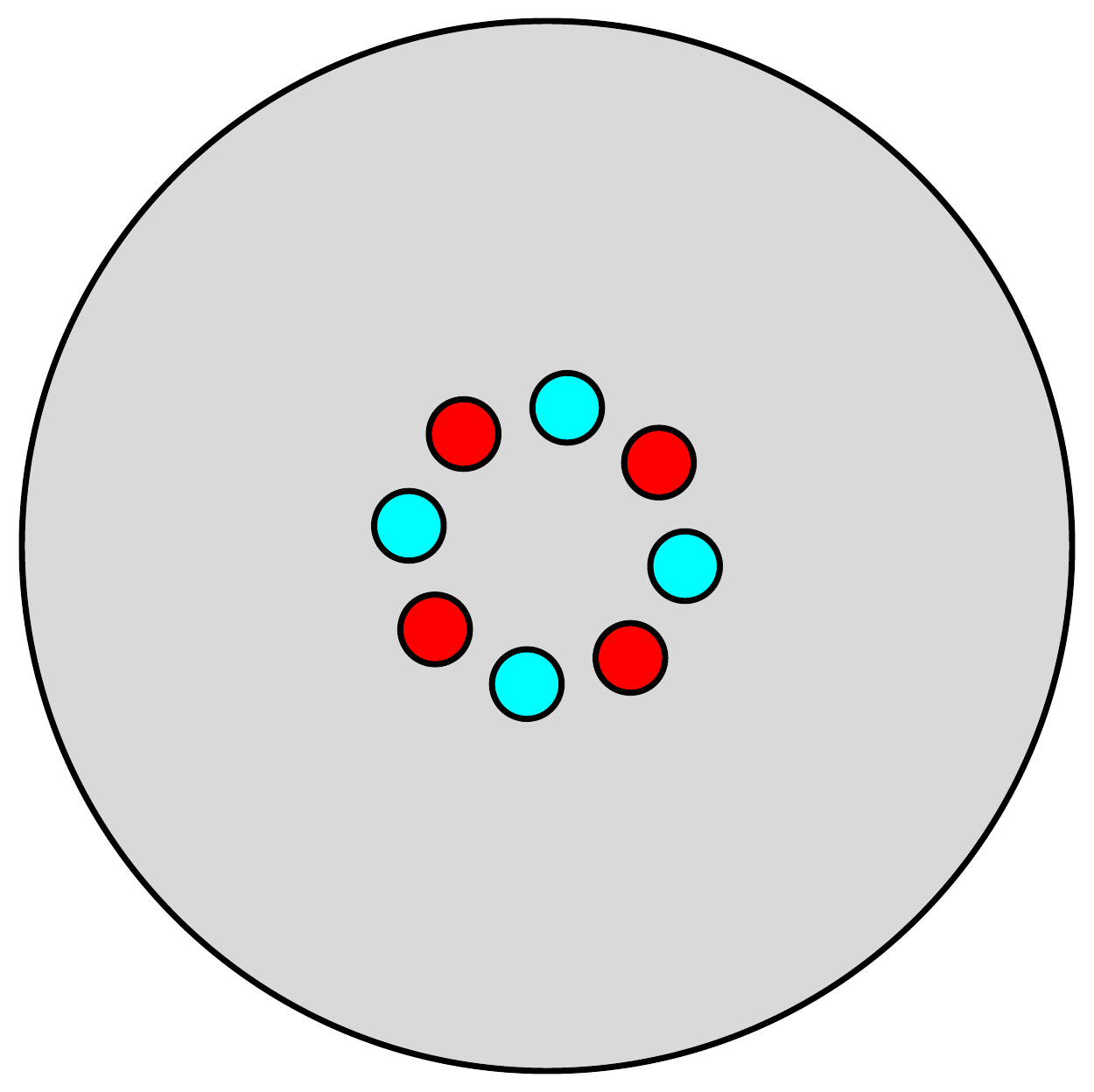}\caption{$B=8$}
	\end{subfigure}
	\hfill\\
	\begin{subfigure}[b]{0.23\textwidth}
		\includegraphics[width=\textwidth]{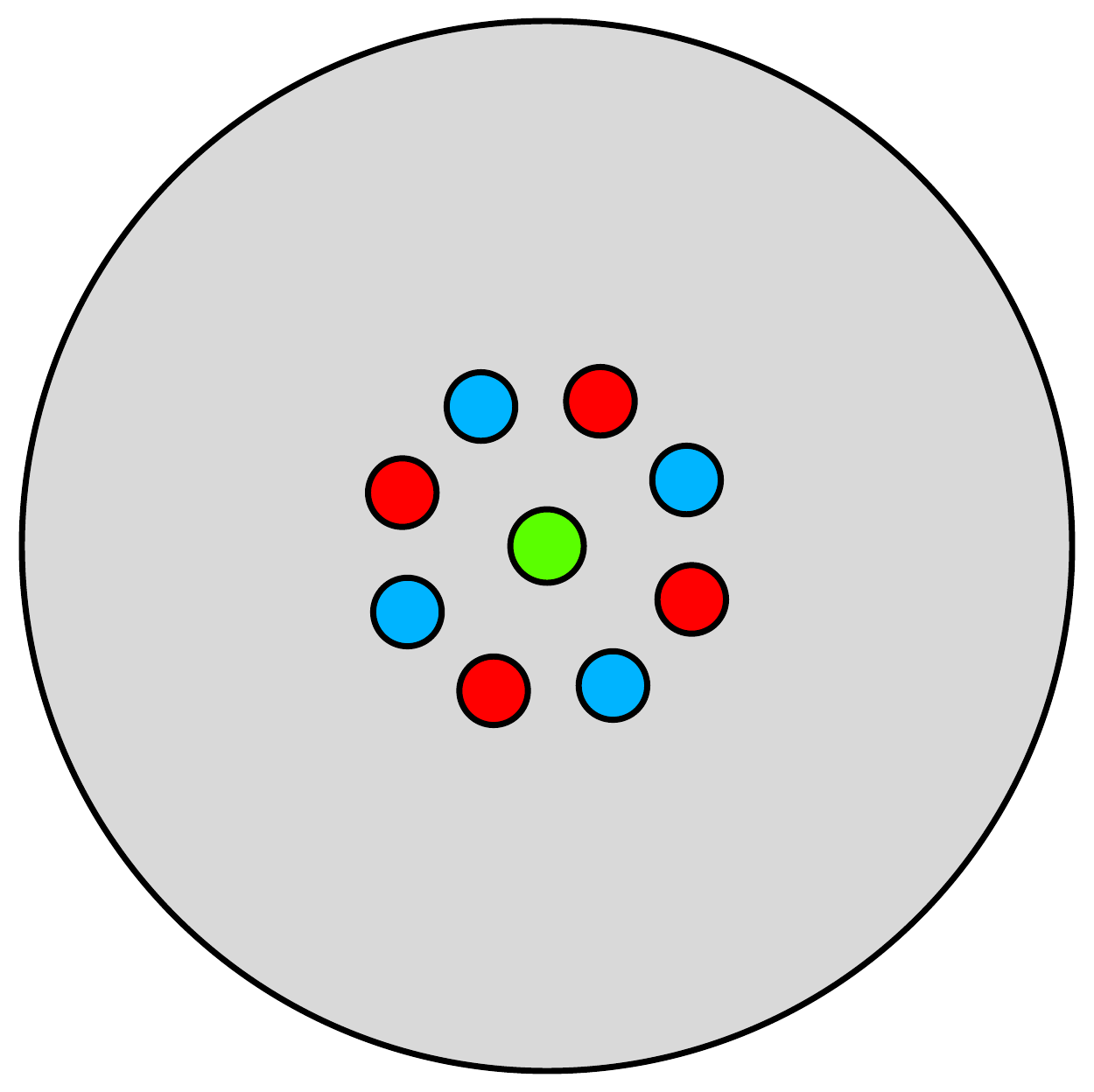}\caption{$B=9$}
	\end{subfigure}
	~
	\begin{subfigure}[b]{0.23\textwidth}
		\includegraphics[width=\textwidth]{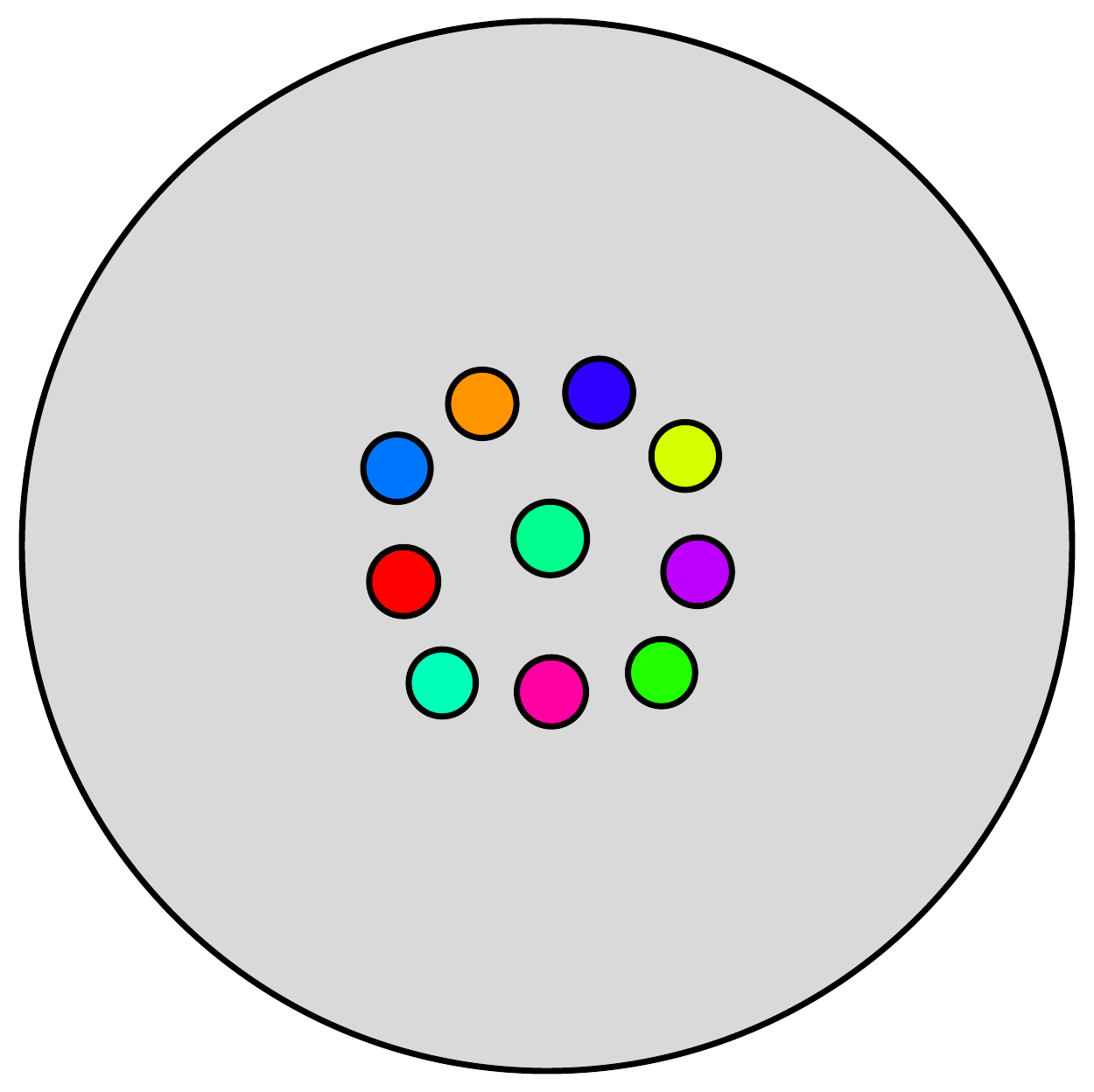}\caption{$B=10$}
	\end{subfigure}
	~
	\begin{subfigure}[b]{0.23\textwidth}
		\includegraphics[width=\textwidth]{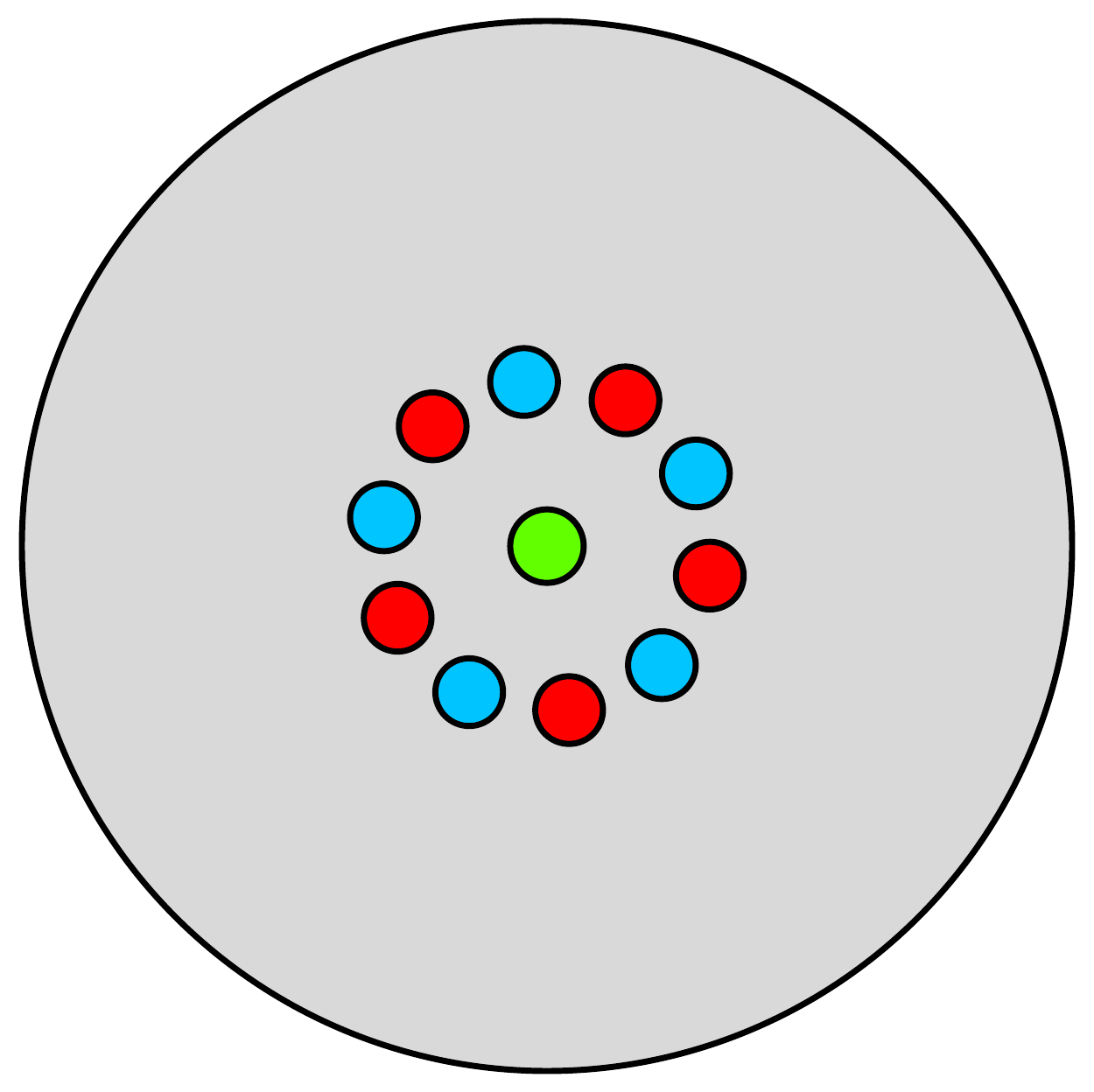}\caption{$B=11$}
	\end{subfigure}
	~
	\begin{subfigure}[b]{0.23\textwidth}
		\includegraphics[width=\textwidth]{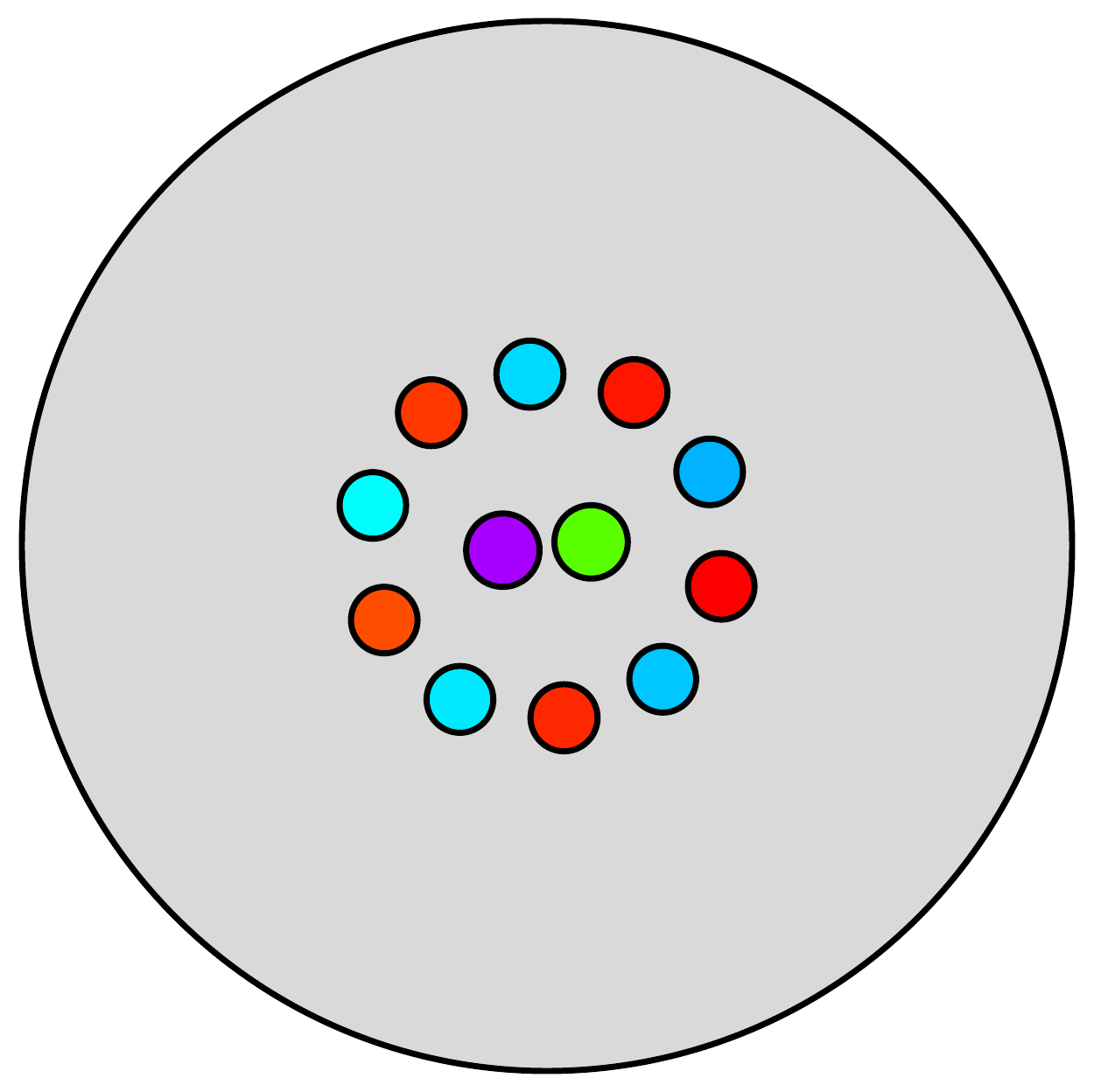}\caption{$B=12$}
	\end{subfigure}
	\hfill\\
	\begin{subfigure}[b]{0.23\textwidth}
		\includegraphics[width=\textwidth]{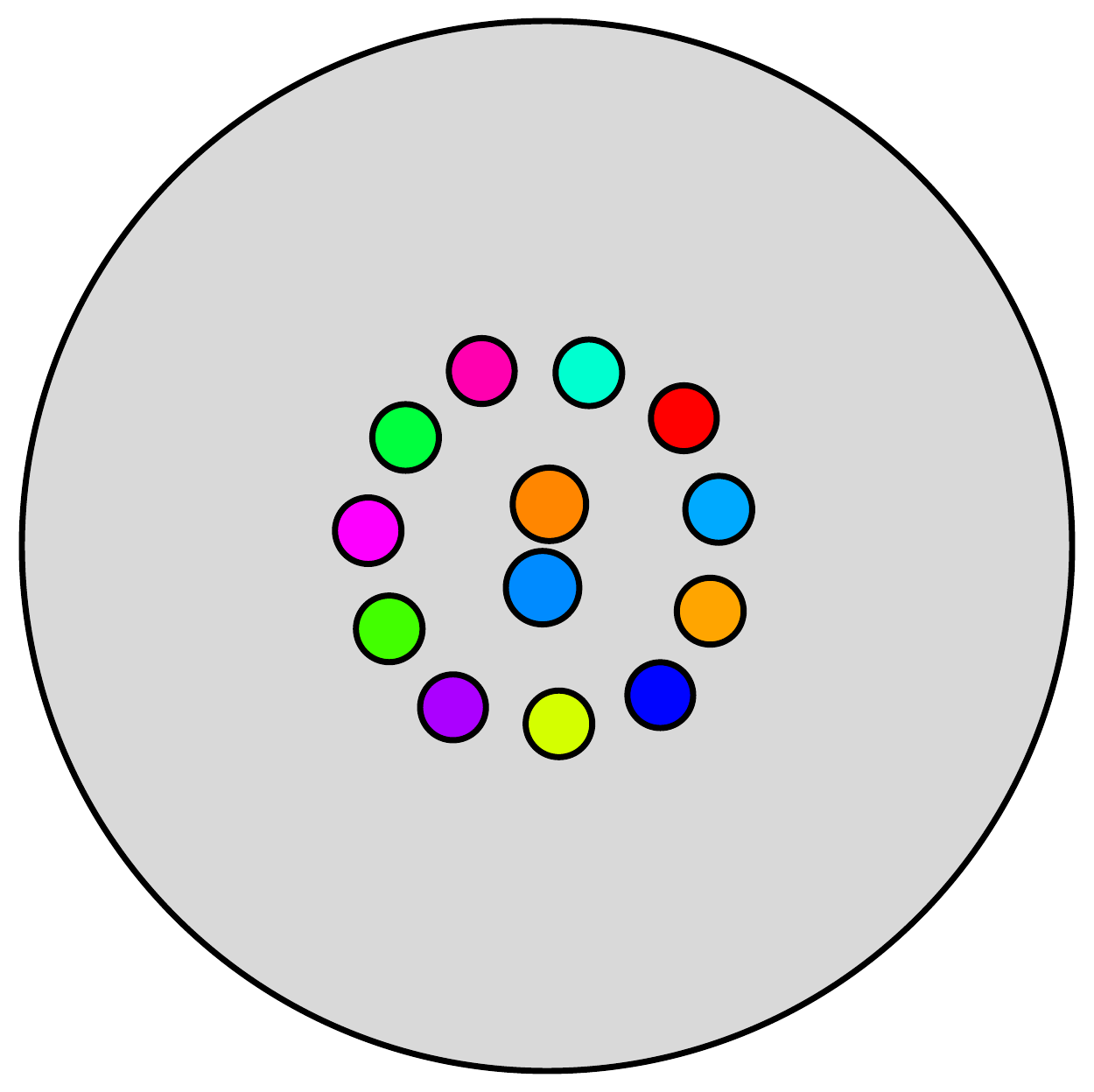}\caption{$B=13$}
	\end{subfigure}
	~
	\begin{subfigure}[b]{0.23\textwidth}
		\includegraphics[width=\textwidth]{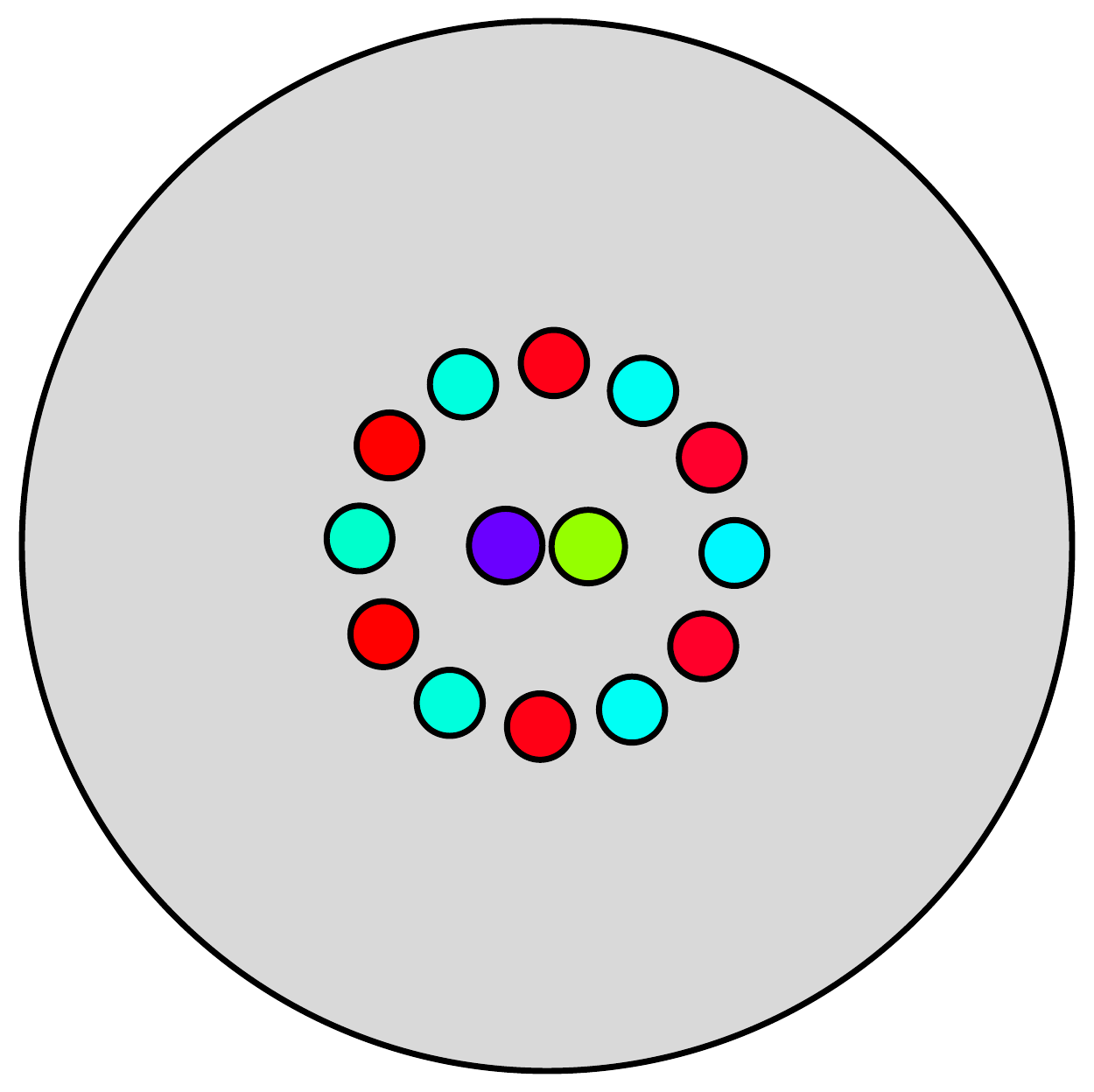}\caption{$B=14$}
	\end{subfigure}
	~
	\begin{subfigure}[b]{0.23\textwidth}
		\includegraphics[width=\textwidth]{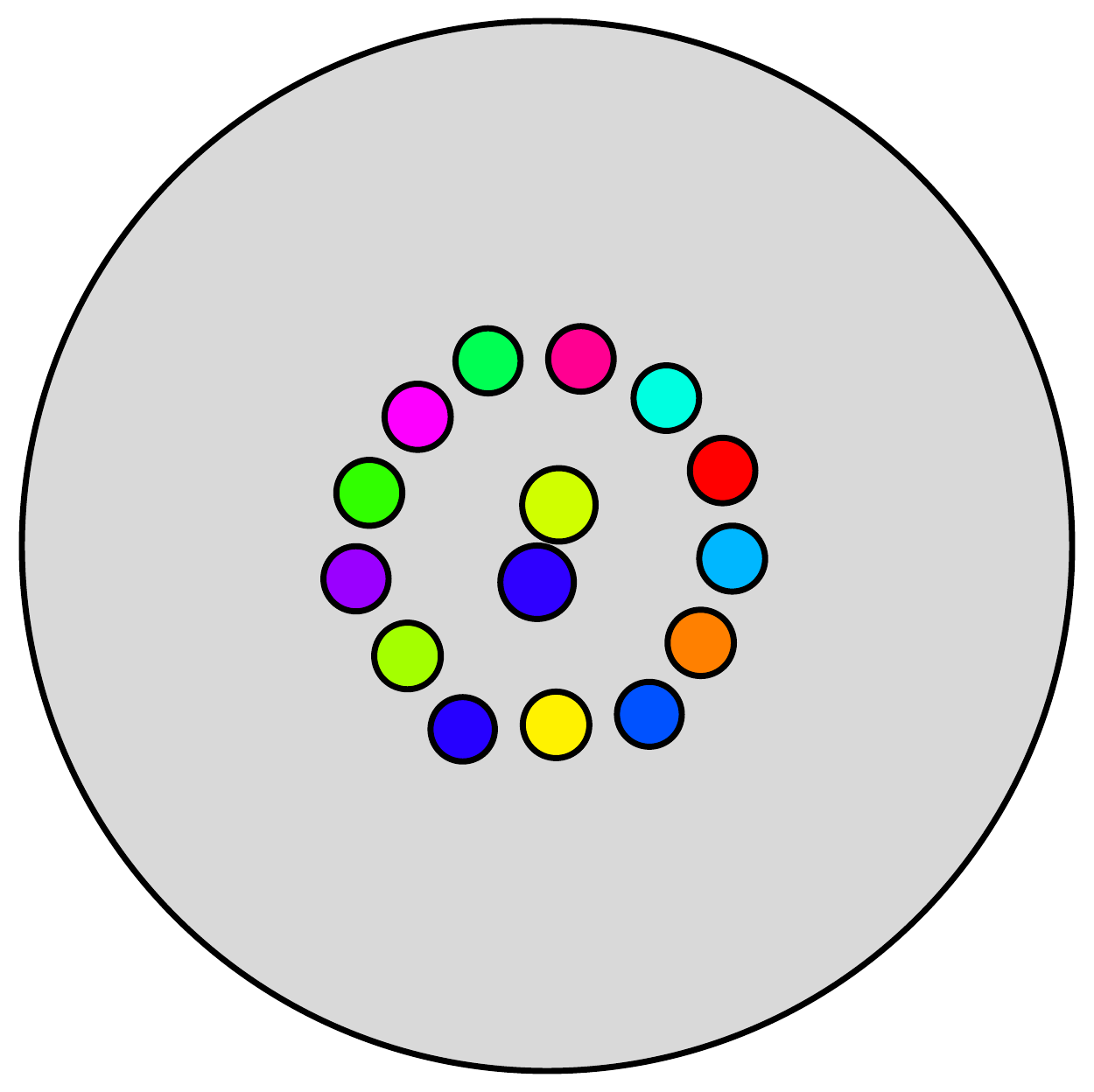}\caption{$B=15$}
	\end{subfigure}
	~
	\begin{subfigure}[b]{0.23\textwidth}
		\includegraphics[width=\textwidth]{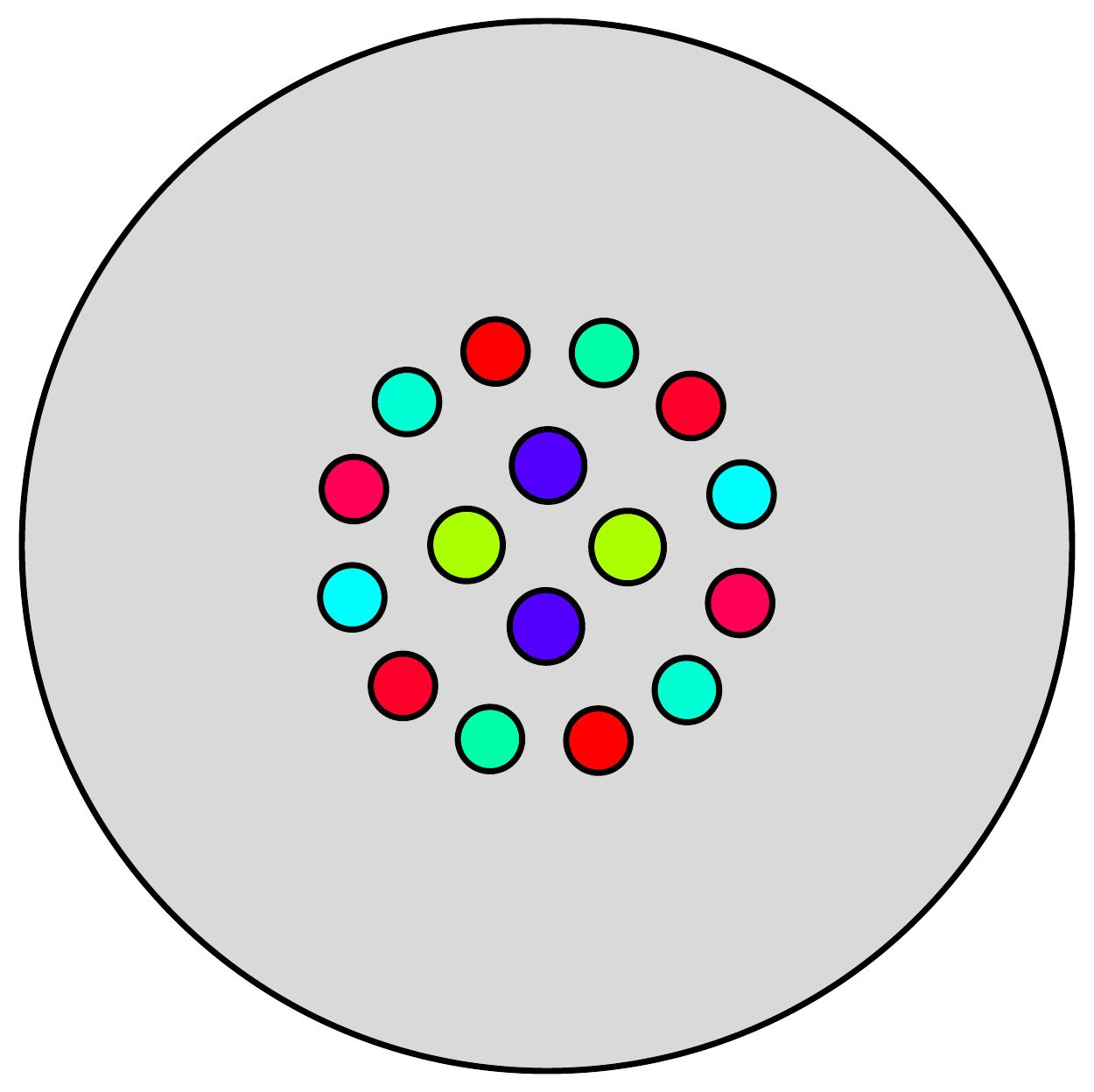}\caption{$B=16$}
	\end{subfigure}
	\hfill\\
	\begin{subfigure}[b]{0.23\textwidth}
		\includegraphics[width=\textwidth]{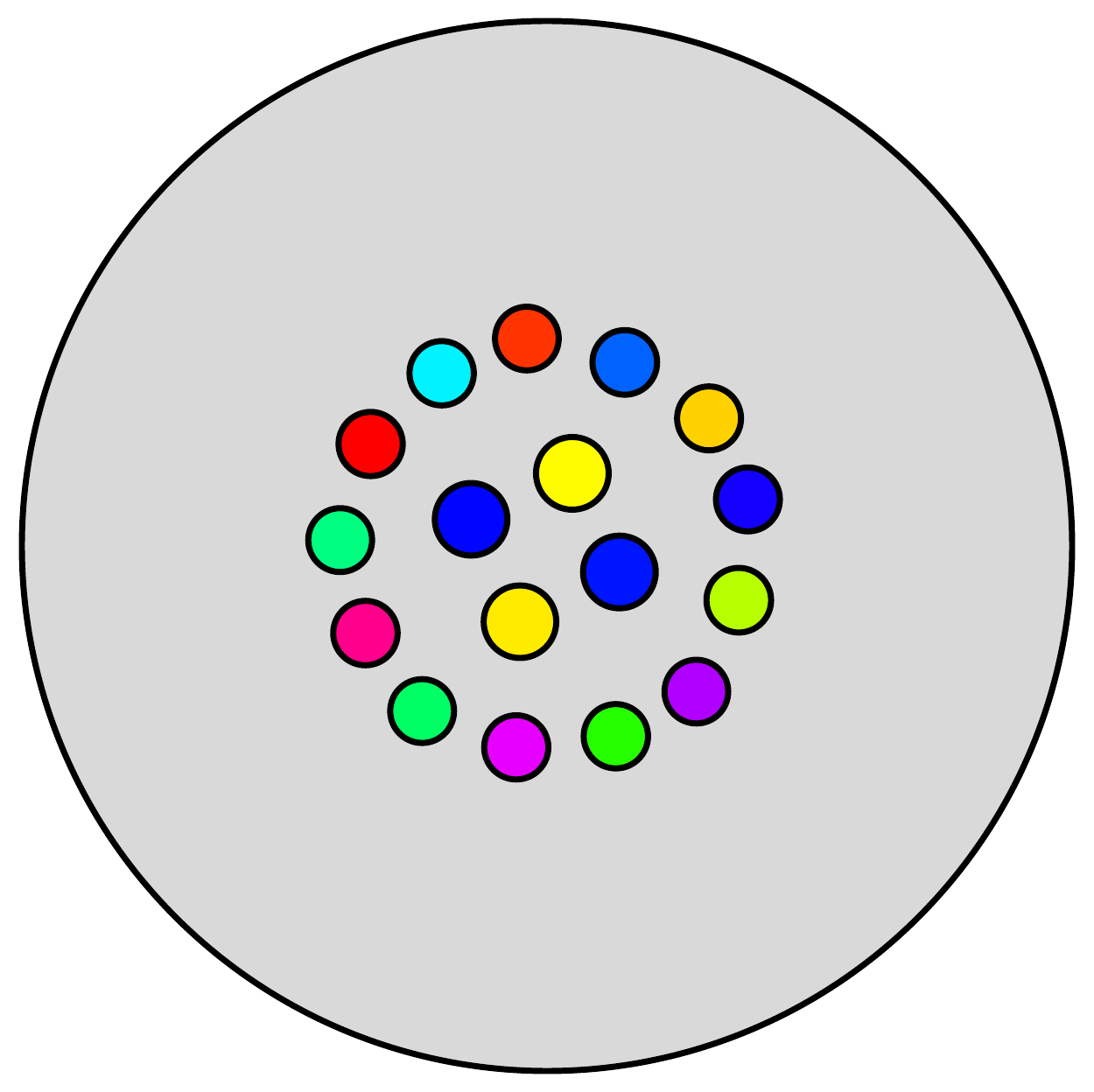}\caption{$B=17$}
	\end{subfigure}
	~
	\begin{subfigure}[b]{0.23\textwidth}
		\includegraphics[width=\textwidth]{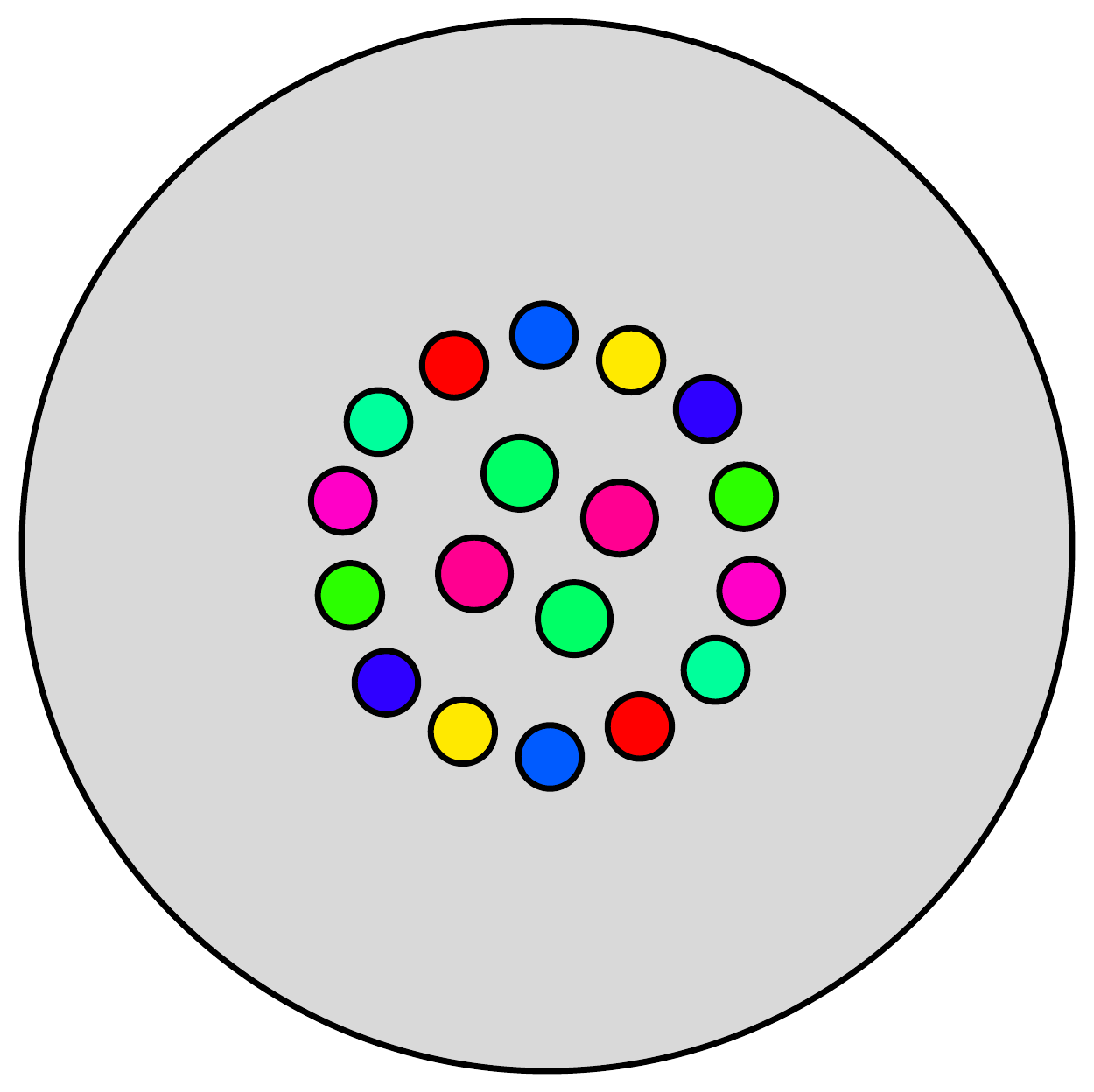}\caption{$B=18$}
	\end{subfigure}
	~
	\begin{subfigure}[b]{0.23\textwidth}
		\includegraphics[width=\textwidth]{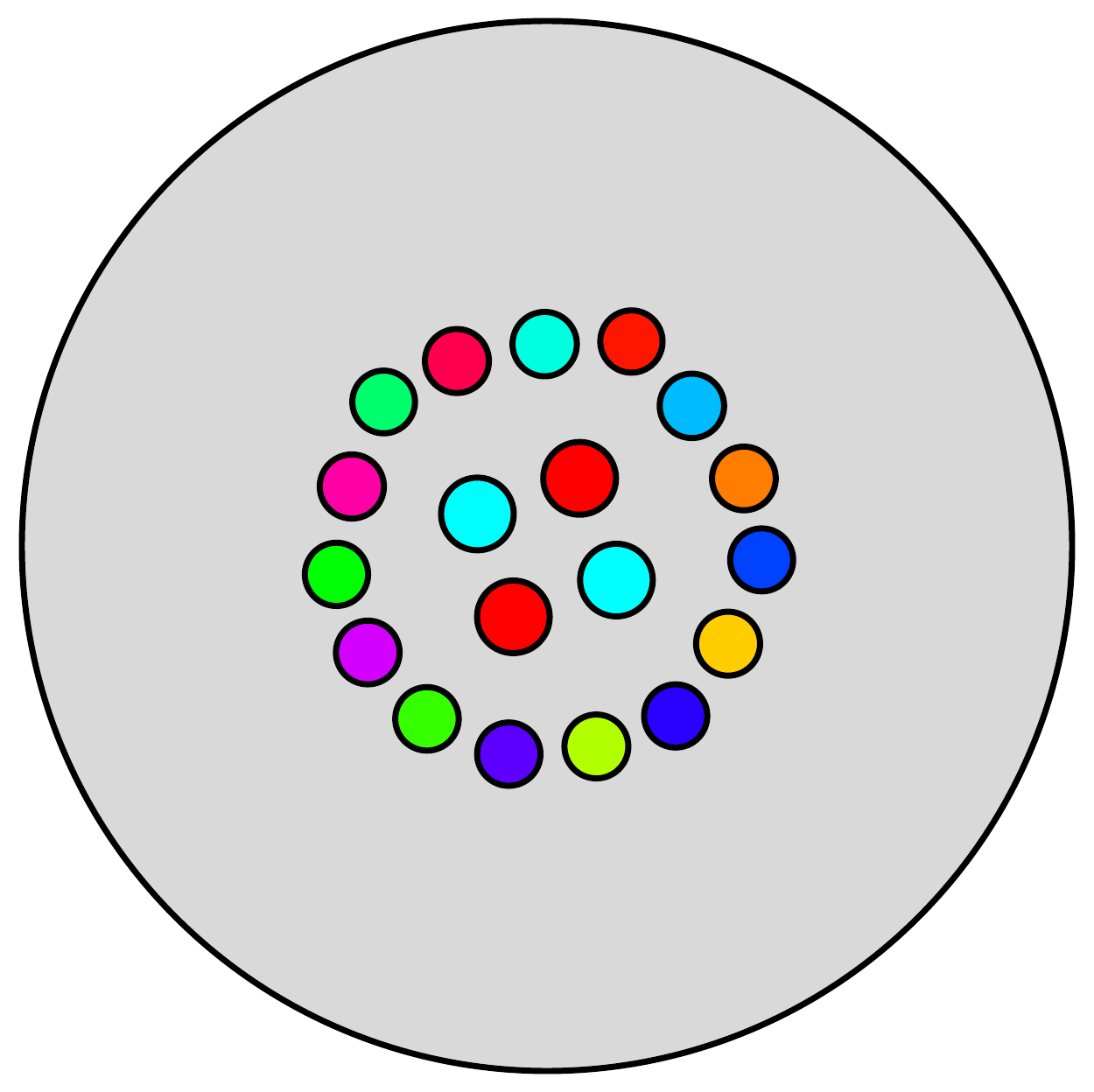}\caption{$B=19$}
	\end{subfigure}
	~
	\begin{subfigure}[b]{0.23\textwidth}
		\includegraphics[width=\textwidth]{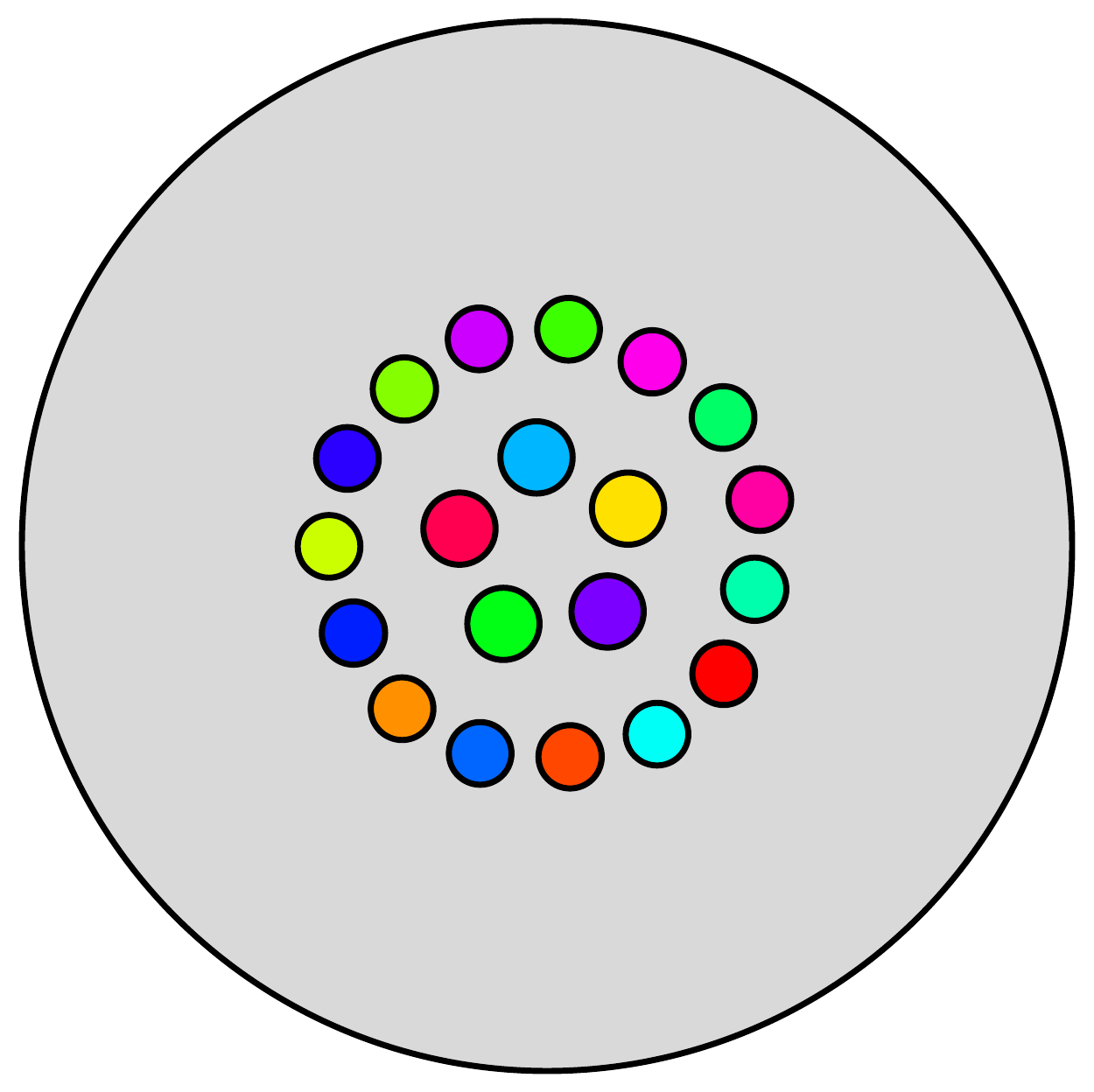}\caption{$B=20$}
	\end{subfigure}
	\hfill\\
	\caption{Minimal energy configurations for the point-particle approximation for topological charges $1\le B\le 20$. Particles are coloured according to their internal phases (see Fig~\ref{phases}).}
	\label{figpoints}
\end{figure}

We minimise the point-particle energy using a multi-start stochastic hill-climbing method, where a randomly generated initial condition is allowed to relax iteratively. For each topological charge we used  randomly generated initial conditions, and the minimum energy configurations found are presented in Fig~\ref{figpoints}. Solutions were verified using a finite temperature annealing method. Particles are coloured according to their phases (see Fig~\ref{phases}).

\begin{figure}[p]
	\centering
	%    \missingfigure[figwidth=8cm]{Graph waiting for numerical simulations to finish}
	\begin{subfigure}[b]{0.23\textwidth}
		\includegraphics[width=\textwidth]{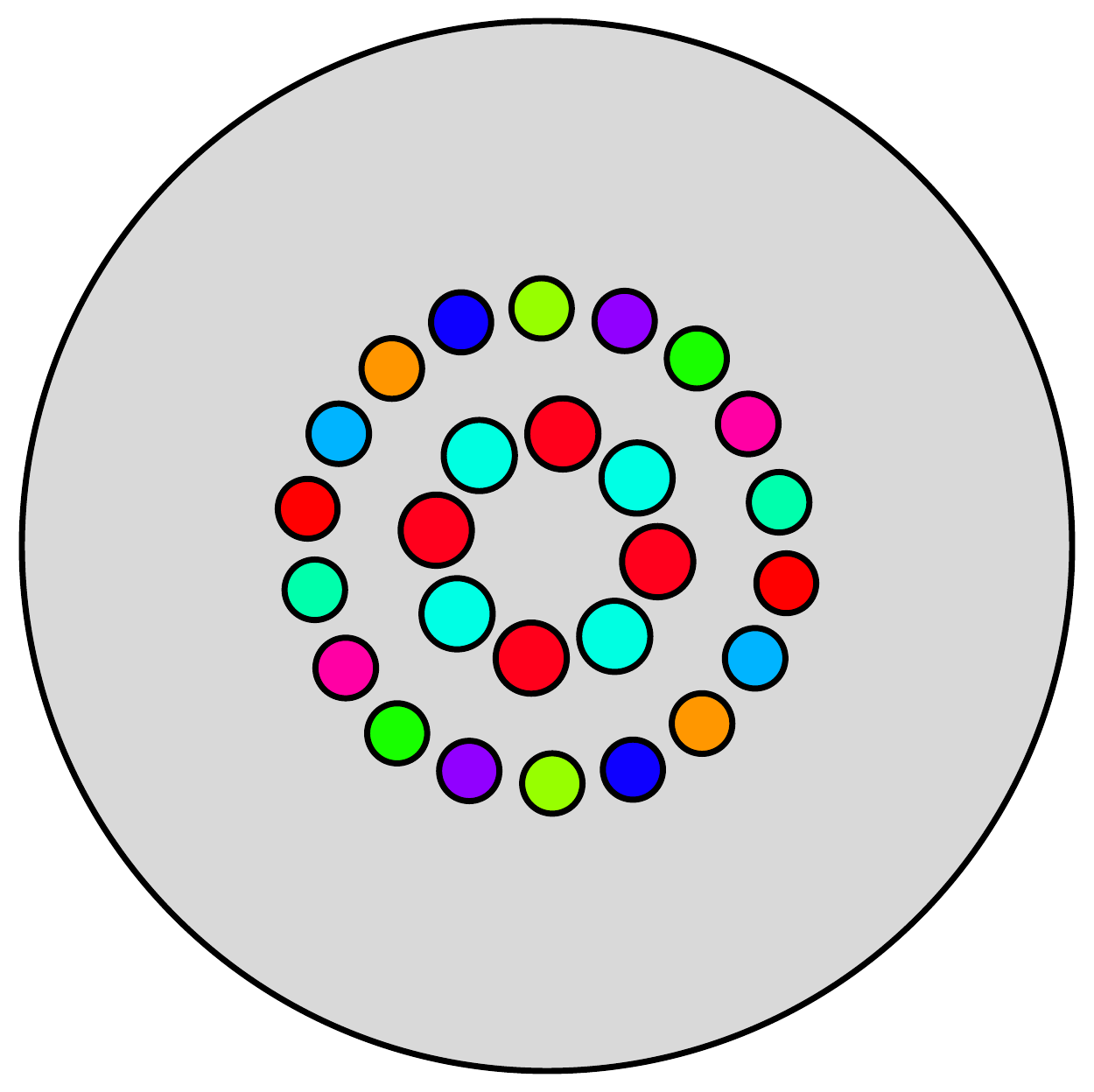}\caption{$B=26$}
	\end{subfigure}
	~
	\begin{subfigure}[b]{0.23\textwidth}
		\includegraphics[width=\textwidth]{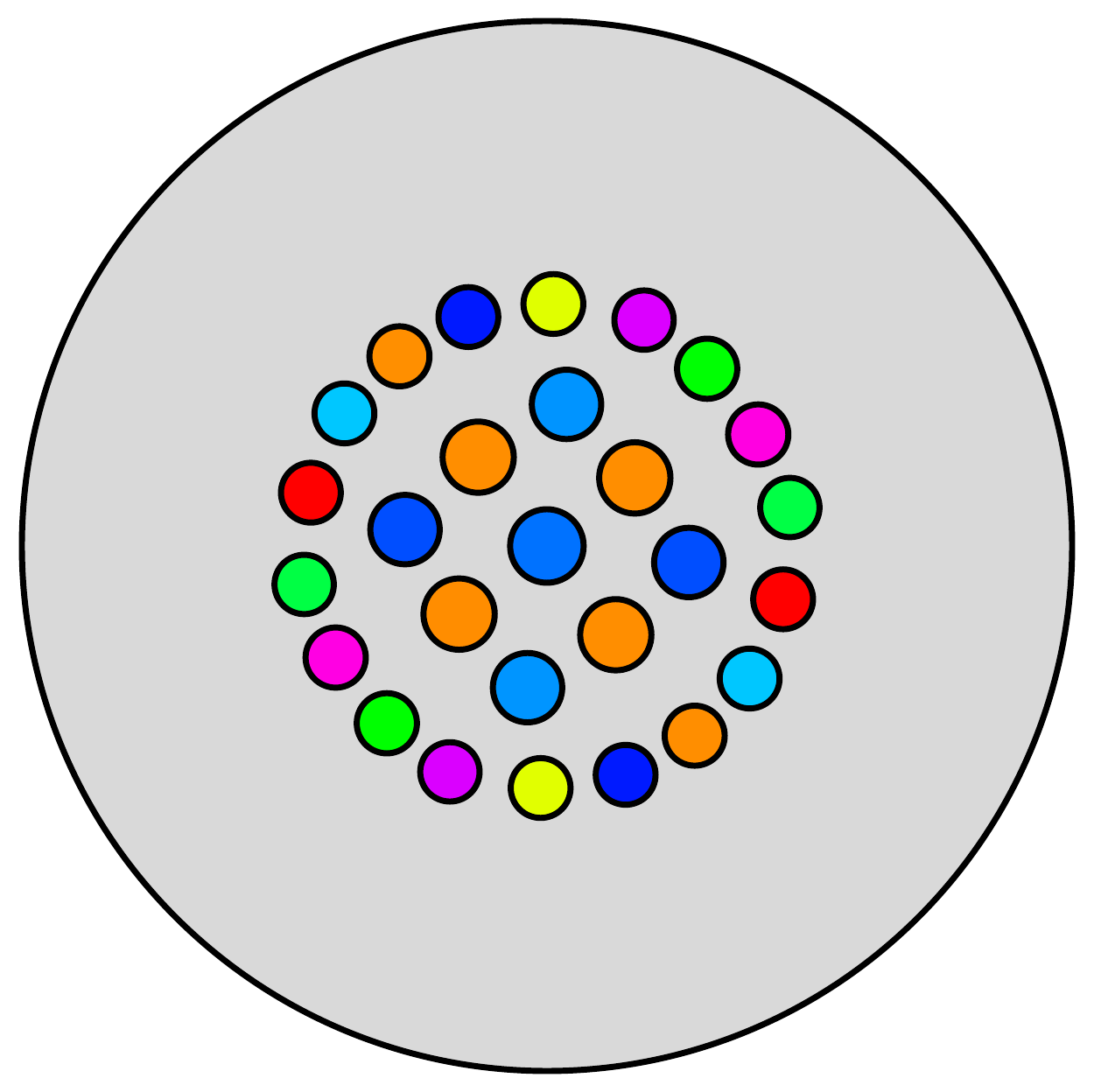}\caption{$B=27$}
	\end{subfigure}
	~
	\begin{subfigure}[b]{0.23\textwidth}
		\includegraphics[width=\textwidth]{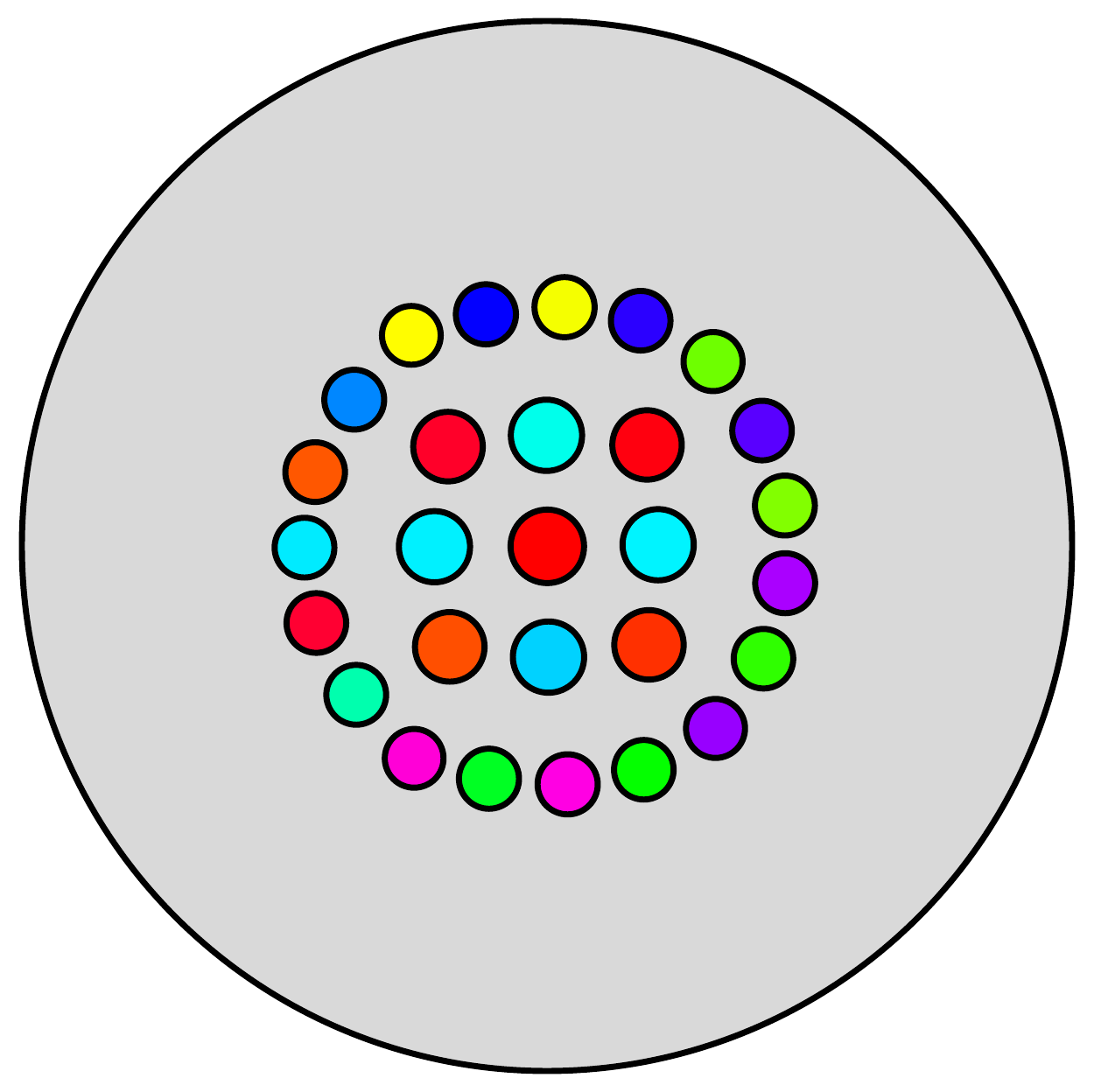}\caption{$B=28$}
	\end{subfigure}
	\hfill\\
	\begin{subfigure}[b]{0.23\textwidth}
		\includegraphics[width=\textwidth]{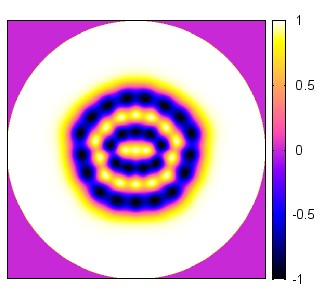}\caption{$B=26$}
	\end{subfigure}
	~
	\begin{subfigure}[b]{0.23\textwidth}
		\includegraphics[width=\textwidth]{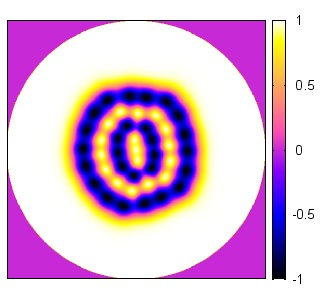}\caption{$B=27$}
	\end{subfigure}
	~
	\begin{subfigure}[b]{0.23\textwidth}
		\includegraphics[width=\textwidth]{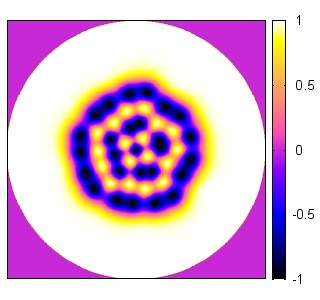}\caption{$B=27*$}
	\end{subfigure}
	~
	\begin{subfigure}[b]{0.23\textwidth}
		\includegraphics[width=\textwidth]{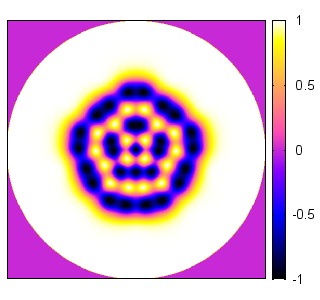}\caption{$B=28$}
	\end{subfigure}
	\hfill\\
	\caption{The top row shows minimal energy configurations for the point-particle approximation for topological charges $26\le B\le 28$. A popcorn transition to a three-layer solution is clearly seen at $B=27$. This is confirmed by full numerical field calculations (below) which demonstrate a popcorn transition around $B=27$, $28$. At $B=27$ two local minima with the same energy (to five significant figures) were found, indicated by the asterisk.}
	\label{pointpops1}
\end{figure}

\begin{figure}[p]
	\centering
	%    \missingfigure[figwidth=8cm]{Graph waiting for numerical simulations to finish}
	\begin{subfigure}[b]{0.20\textwidth}
		\includegraphics[width=\textwidth]{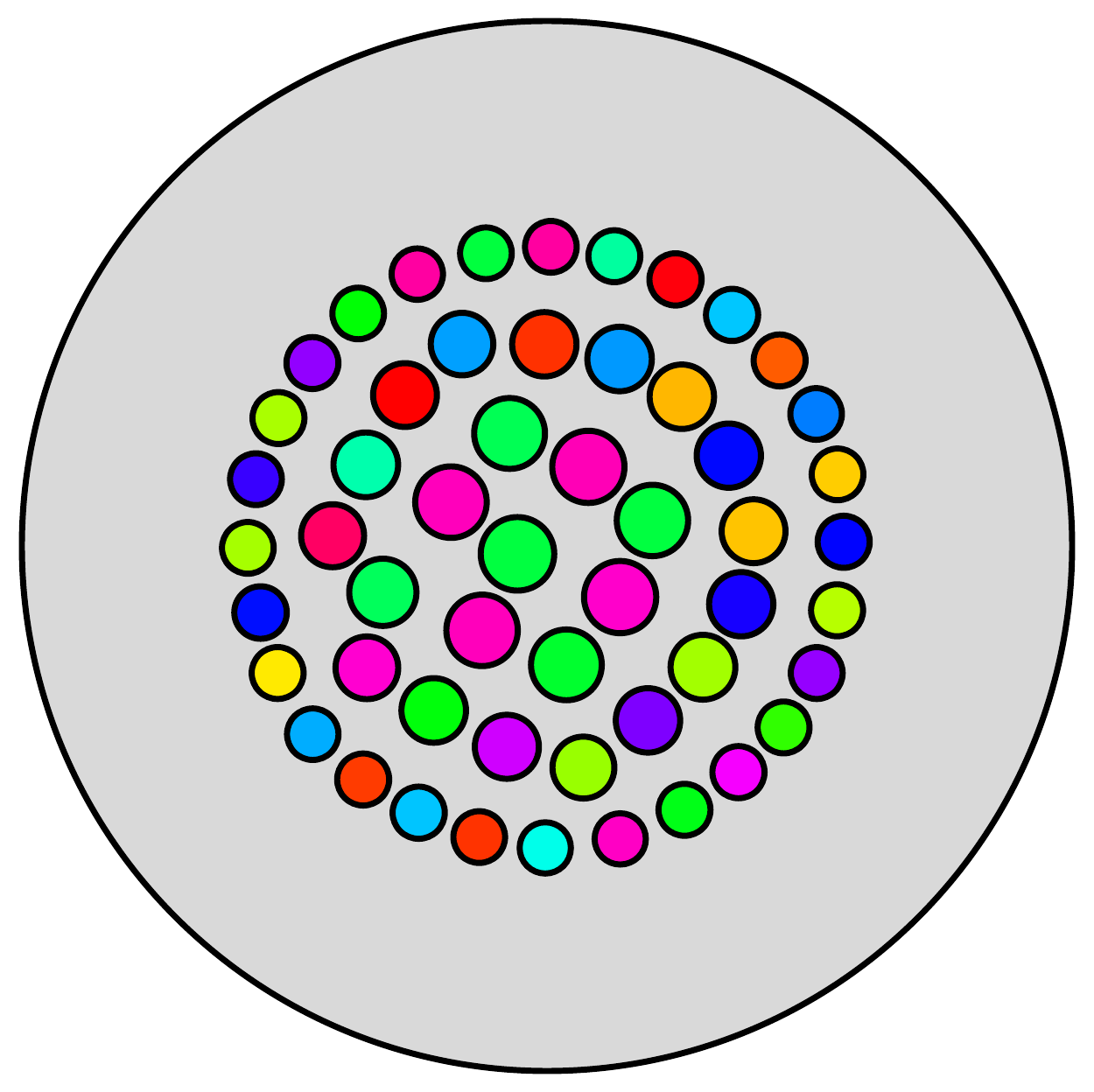}\caption{$B=53$}
	\end{subfigure}
	~
	\begin{subfigure}[b]{0.20\textwidth}
		\includegraphics[width=\textwidth]{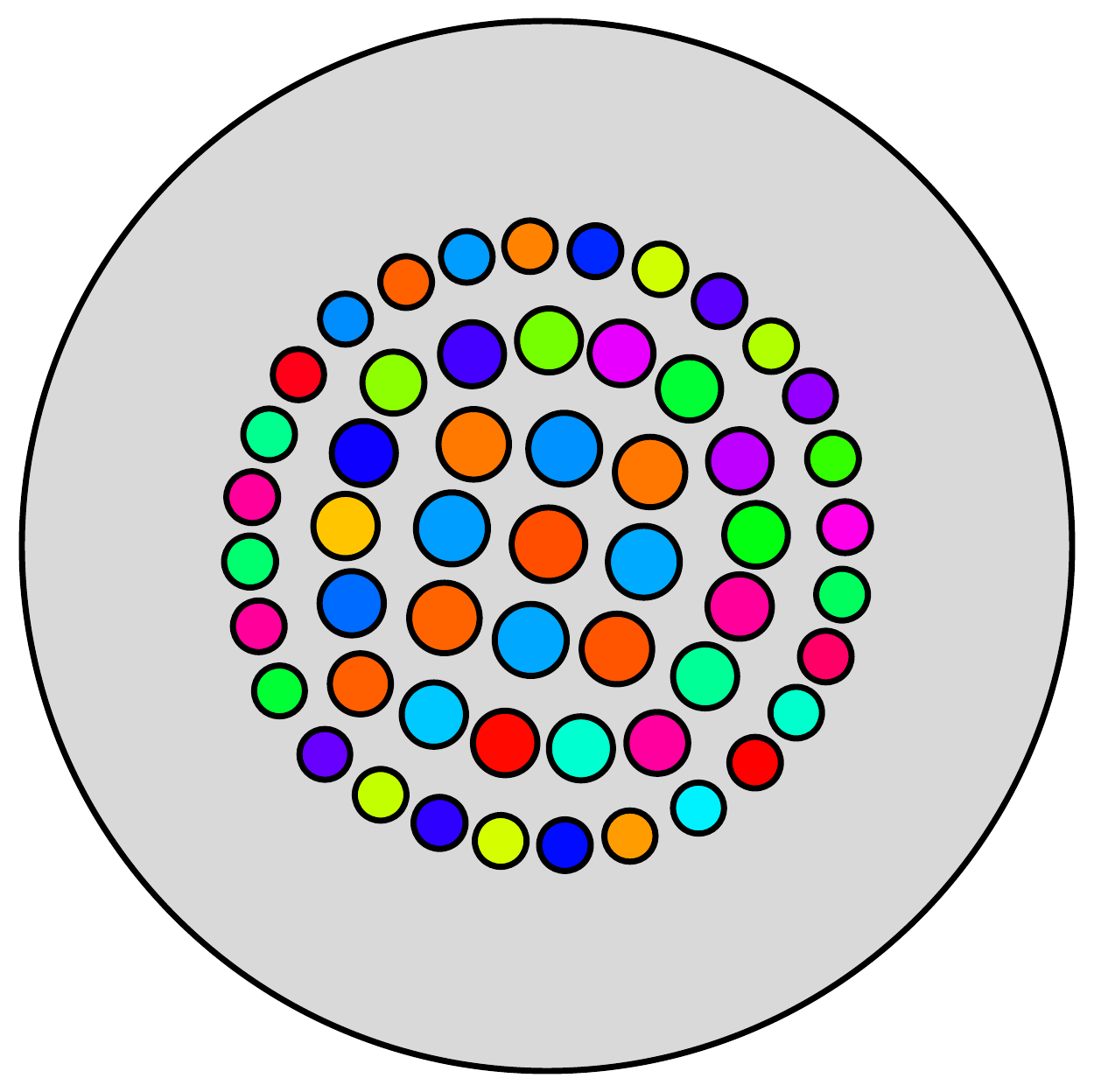}\caption{$B=54$}
	\end{subfigure}
	\hfill\\
	\begin{subfigure}[b]{0.20\textwidth}
		\includegraphics[width=\textwidth]{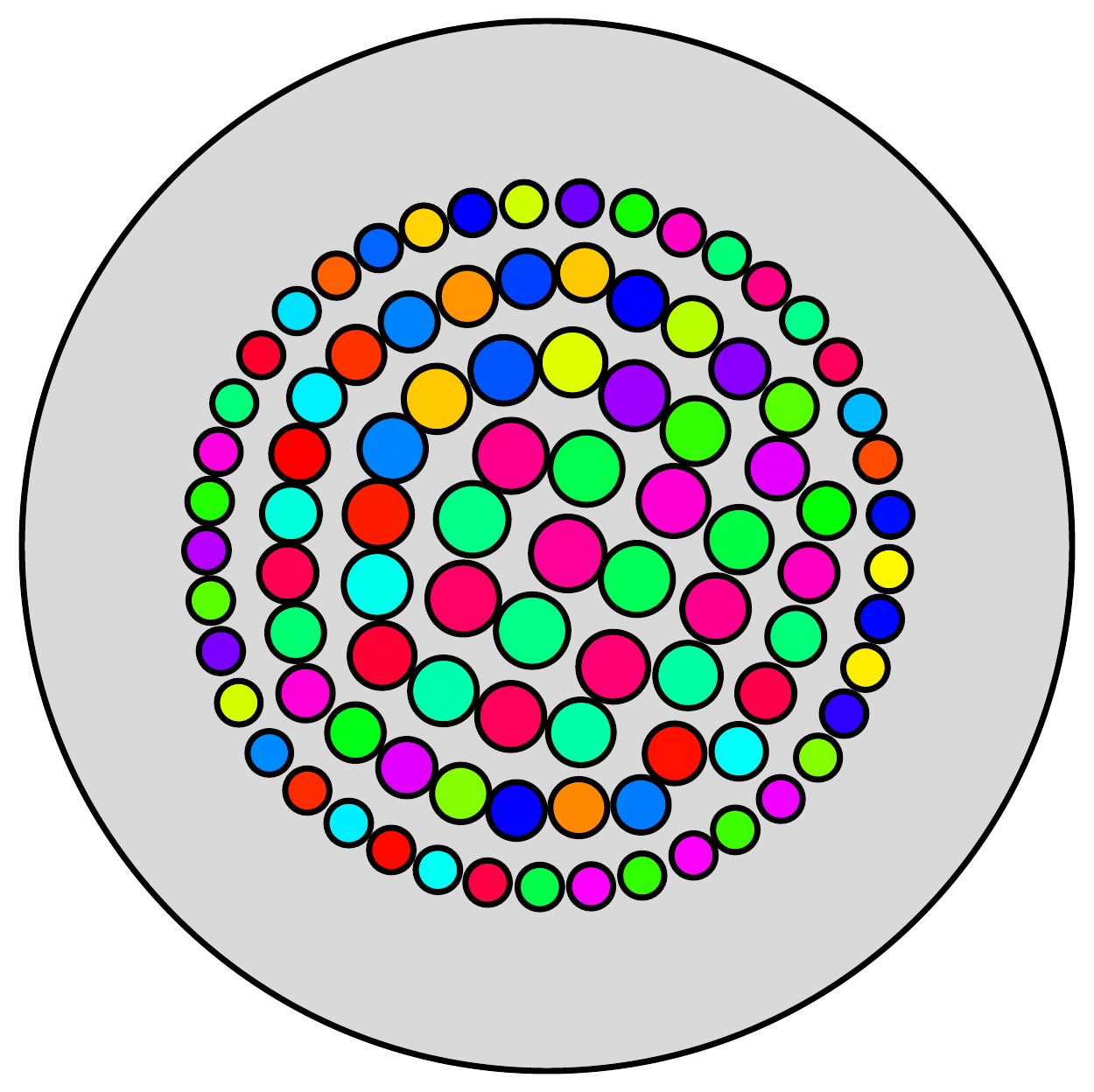}\caption{$B=93$}
	\end{subfigure}
	~
	\begin{subfigure}[b]{0.20\textwidth}
		\includegraphics[width=\textwidth]{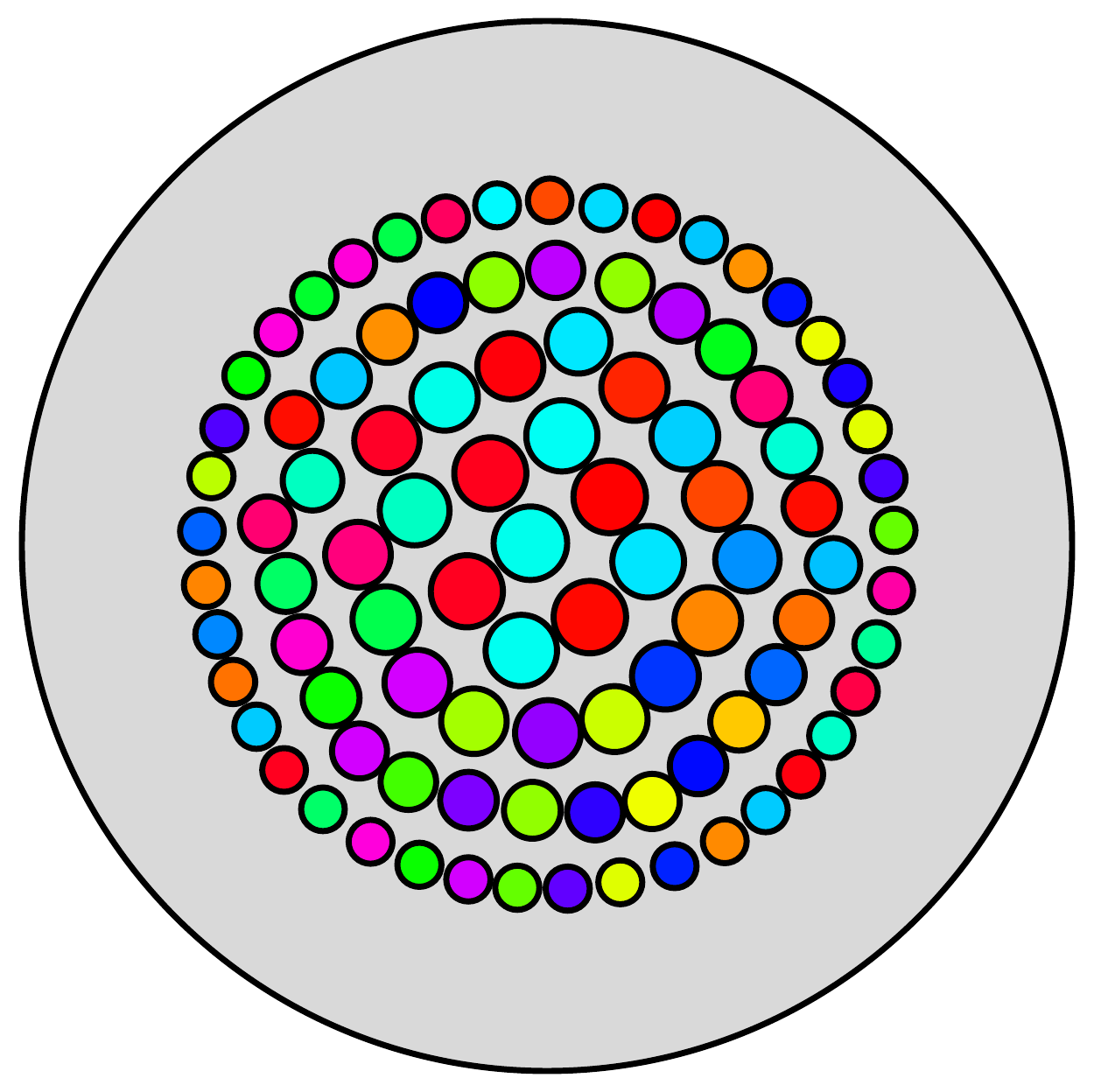}\caption{$B=94$}
	\end{subfigure}
	~
	\begin{subfigure}[b]{0.20\textwidth}
		\includegraphics[width=\textwidth]{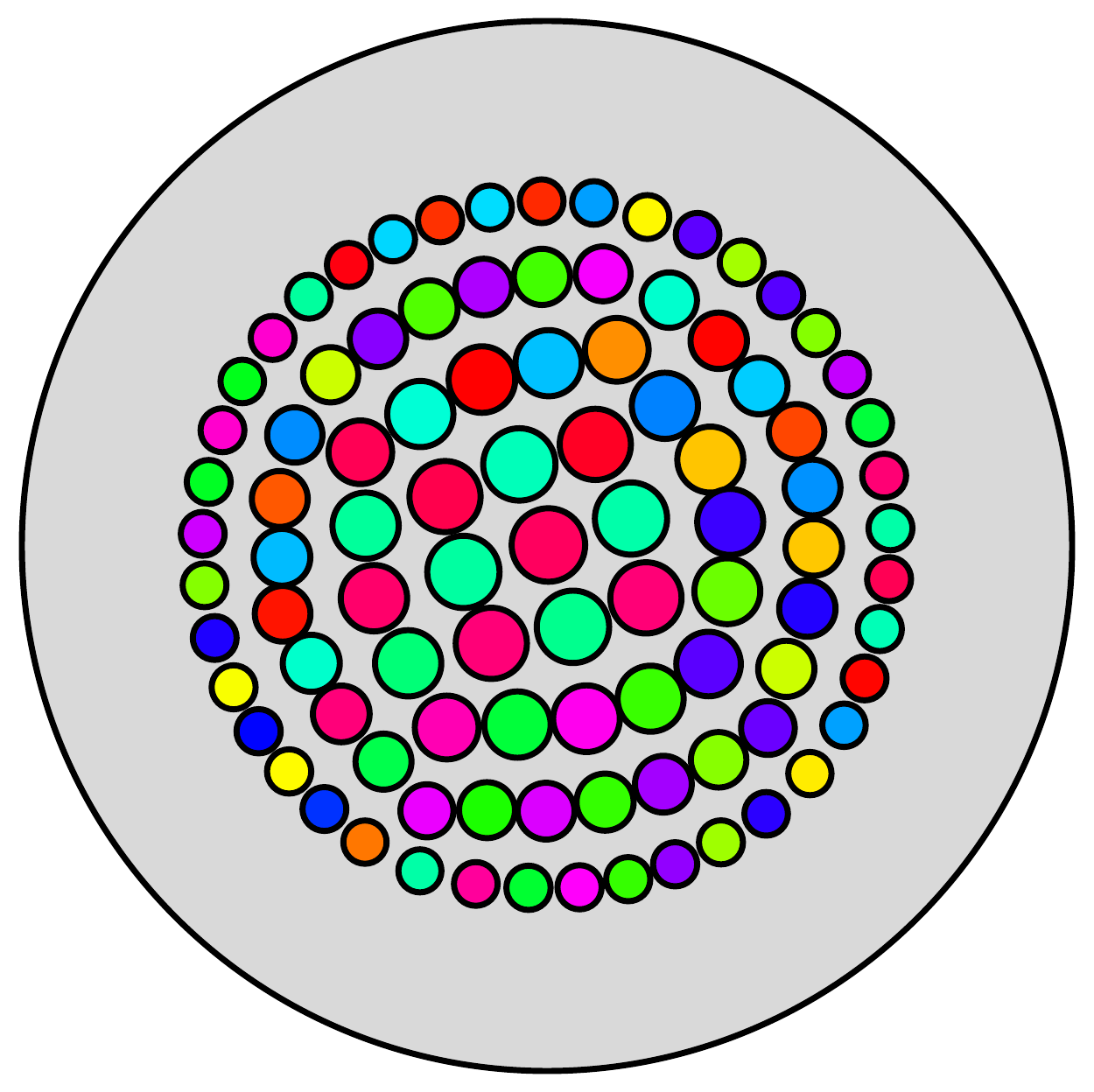}\caption{$B=95$}
	\end{subfigure}
	\hfill\\
	\caption{Minimal energy configurations of the point-particle approximation demonstrating popcorn transitions to four and five layers at charges $B=54$ and $B=95$ respectively. The lower charges pictured show how the inner-most ring of each configuration becomes deformed just before a pop.}
	\label{higherpops}
\end{figure}

We find that the point-particle approximation we have derived also favours configurations that form concentric ring-like structures. Furthermore, the qualitative forms of the energy minima predicted by the point-particle approximation are very close to the forms of the full numerical solutions found in Fig~\ref{adsbabies}, although the results of the point-particle approximation are much more symmetric. This is to be expected since solitons in the full model have a finite size, and so can overlap and interact in more complicated ways.

For charges $B\le 7$ the approximation accurately predicts not only the ring structure of the configurations, but also the alignment of internal phases. For even and odd charges the particles have internal phase differences of $\pi$ and $\pi\pm\pi/B$ respectively, as observed in the full model. At $B=8$ the approximation gives the correct distribution of phases, although it does not reproduce the deformed ring structure.

Furthermore, the point-particle approximation correctly captures the first popcorn transition at $B=9$, and closely estimates the qualitative forms of the ring structures for all $B\le 20$. The point-particle minima disagree with the full numerical results at charges $B=11$, $13$, $15$, $16$, $19$ and $20$. This may be due to the point-particle approximation not taking into consideration the radial forms of the $B=2$ and $3$ solitons, or may be a result of assuming the solitons can be approximated by particles with zero size. However, even when the exact forms do not agree, the difference is only by the position of a single particle.

These results suggest that the point-particle approximation may be a useful tool in qualitatively estimating the forms taken by AdS baby Skyrmions for higher charges. Performing further numerical minimisations of \eqref{ptptclenergy} allows us to predict a second popcorn transition at charge $B=27$. In fact, by using the predicted forms as a guide for choosing initial conditions, full numerical energy minimisations reveal the popcorn transition to three layers around charge $B=27$, $28$, although it is difficult to say exactly when the transition occurs due to the presence of two local minima with similar energies (shown in Fig~\ref{pointpops1}, with energies given in Table~\ref{popstable}).

\begin{table}[t]
	\centering
	\begin{tabular}{ | c c c | }
		\hline
		charge $B$ & $E/4\pi B$ & form \\
		\hline                       
		26 & 1.8357 & \{9, 17\} \\
		27 & 1.8546 & \{9, 18\}  \\
		27* & 1.8546 & \{1, 9, 17\} \\	
		28 & 1.8723 & \{1, 9, 18\}  \\
		
		\hline
	\end{tabular}
	\caption{Energies per soliton and forms for AdS baby Skyrmions with topological charge $26\le B\le 28$. Two local minima of the same energy (to five significant figures) were found at $B=27$, as indicated by the asterisk.}
	\label{popstable}
\end{table}

\begin{figure}[b]
	\centering
	%    \missingfigure[figwidth=8cm]{Graph waiting for numerical simulations to finish}
%	\includegraphics[width=0.23\textwidth]{./PtPtcles/approx_B200.pdf}
	\begin{subfigure}[b]{0.20\textwidth}
		\includegraphics[width=\textwidth]{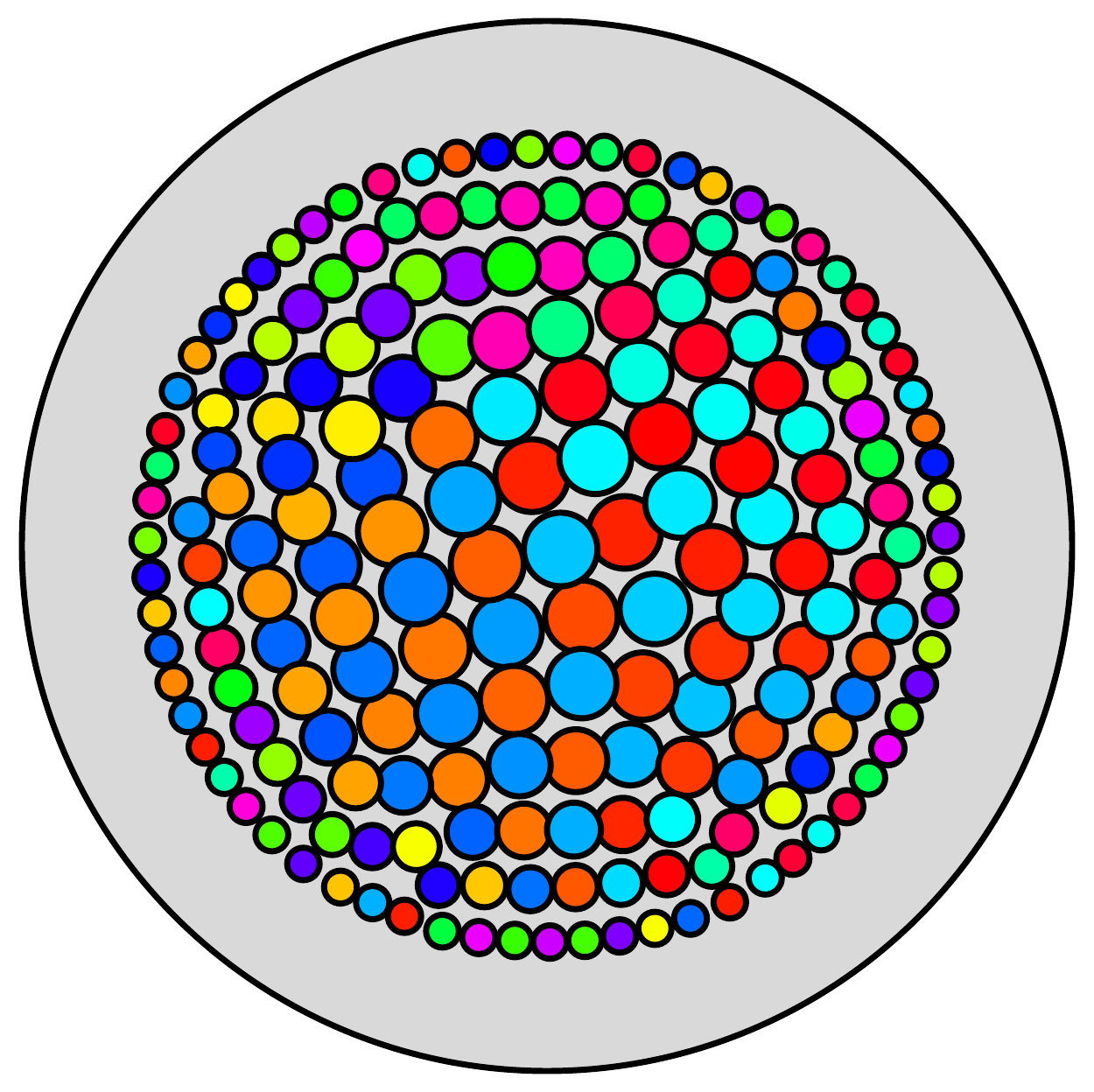}\caption{$B=200$}
	\end{subfigure}
	~
	\begin{subfigure}[b]{0.20\textwidth}
		\includegraphics[width=\textwidth]{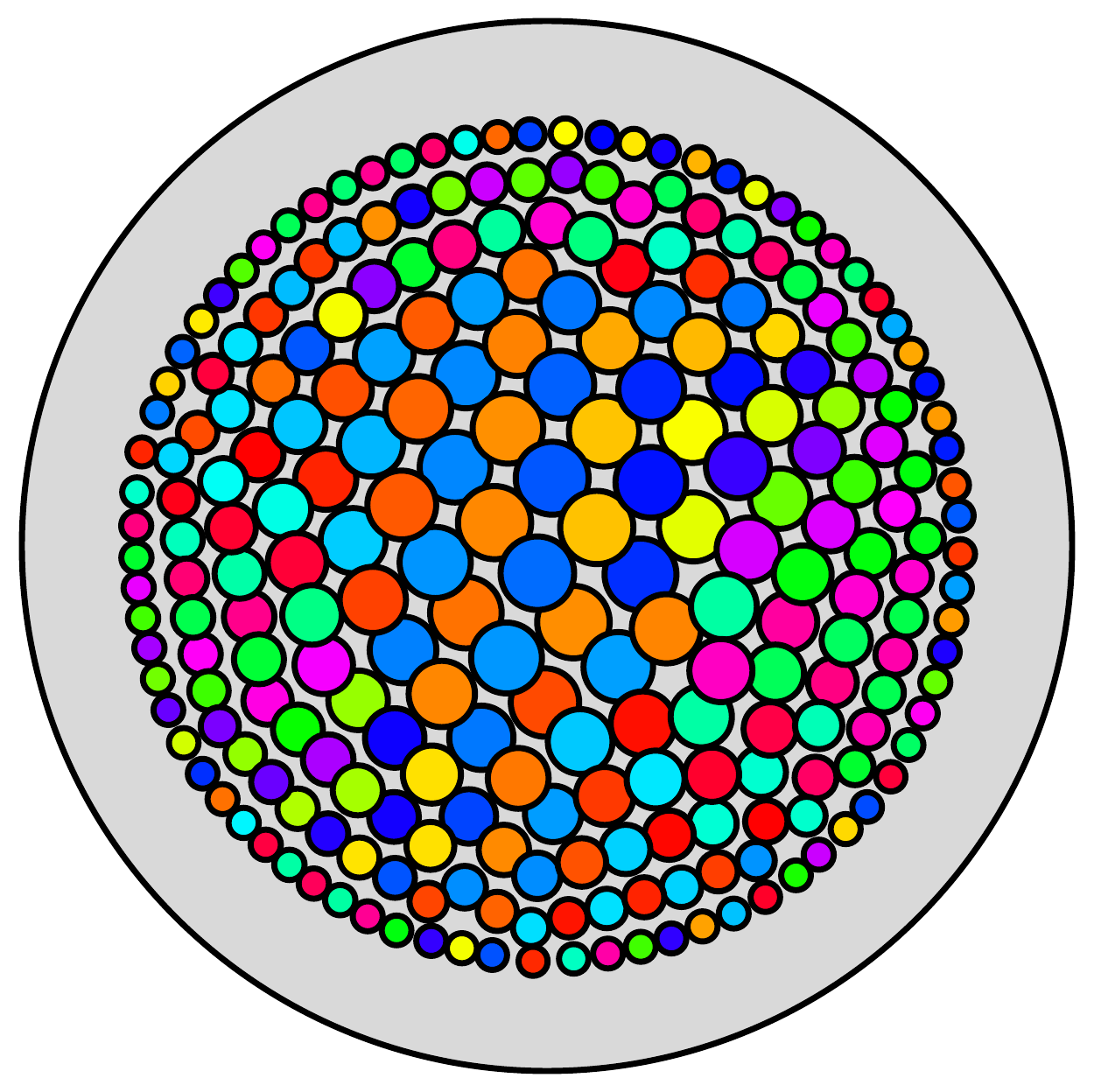}\caption{$B=250$}
	\end{subfigure}
	\hfill
	\caption{The minimal energy configuration of the point-particle approximation at charges $B=200$ and $250$. The ring structure near the origin is no longer visible, and may indicate the emergence of a lattice structure.}
	\label{pointlarge}
\end{figure}

The point-particle approximation predicts a third and fourth popcorn transition at charges $B=54$ and $B=95$ (see Fig~\ref{higherpops}). Since these charges are very large it would be difficult to perform an extensive search for the global minima in the full numerical model, even with the guidance provided by the point-particle approximation. However, the previous success of the model would indicate that the true popcorn transitions would, indeed, be near these points.

Finally, investigation of the results predicted by the point-particle approximation for still higher charges may be able to provide clues as to the lattice structure preferred by the baby Skyrmions in the infinite charge limit. The energy minimum for charge $B=200$ can be seen in Fig~\ref{pointlarge}. While clear rings can still be observed in the outermost layers, the centre of the configuration is heavily deformed and may indicate an emerging lattice structure.

\section{Conclusions}
We have investigated the static solitons and multi-solitons of the massless baby Skyrme model in a $(2+1)$-dimensional Anti de-Sitter spacetime. We have found that the spacetime curvature acts by adding an effective mass to the model which allows us to find static solutions to the equations of motion without a mass term. Solitons for topological charges $1\le B\le 3$ were found to have a radially symmetric form, while higher-charge multi-solitons were found to form concentric ring-like solutions.

As $B$ increases, a series of transitions occur where the minimal energy solutions take the form of concentric rings with increasing numbers of layers, a phenomenon reminiscent of the baryonic popcorn transitions studied recently in the context of holographic dense QCD. In order to investigate these transitions further a point-particle approximation was derived which was able to qualitatively estimate the forms of minimal energy solutions for a wide range of topological charges, as well as accurately predict the charges at which further popcorn transitions occur.

The point-particle approximation derived may indicate an emerging lattice structure for baby Skyrmions in AdS in the limit $B\to\infty$. The minimum energy configuration found for $B=200$ displayed clear rings towards the edge of the space, although near the origin the structure appeared significantly less ring-like, and seemed suggestive of an emergent lattice formation. Further investigation into this area would be required to make any further claims.

The $O(3)$-sigma model stabilised by a baby Skyrme term has previously been studied in an AdS-like spacetime as a low-dimensional toy model of holographic QCD \cite{Bolognesi:2013jba}, specifically the Sakai-Sugimoto model. It has been argued that a better low-dimensional analogue of this model would involve stabilising an $O(3)$-sigma model using a vector meson term, and such a model has been studied in a parameter regime where the two toy models are similar \cite{Elliot-Ripley:2015cma}, although investigation of a more interesting parameter regime proved difficult. It would therefore be interesting to study the vector meson model in pure AdS to see if a parameter regime could be found to give qualitatively different results to the baby Skyrme model.

Finally, the natural extension to this paper is to study the full $(3+1)$-dimensional Skyrme model in AdS. We have demonstrated that a multi-ring like structure exists in two dimensions for AdS baby Skyrmions and this property may translate to the higher dimensional model. This could manifest as spherical multi-shells with polyhedral symmetry groups and hence be approximated by multi-shell rational maps and it would be very interesting to study AdS Skyrmions to see if such configurations give lower energy solutions.

\section{Acknowledgements}
Matthew Elliot-Ripley would like to thank STFC and Thomas Winyard EPSRC for funding during our PhD studies. We would also like to thank our supervisor, Paul Sutcliffe, for useful discussions and support.

\bibliography{../../Papers/master.bib}

\providecommand{\href}[2]{#2}\begingroup\raggedright\begin{thebibliography}{10}

\bibitem{Skyrme:1961vq}
T.~Skyrme, {\it {A nonlinear field theory}},  {\em Proc.Roy.Soc.Lond.} {\bf
  A260} (1961) 127.

\bibitem{Sakai:2004cn}
T.~Sakai and S.~Sugimoto, {\it {Low energy hadron physics in holographic QCD}},
   {\em Prog.Theor.Phys.} {\bf 113} (2005) 843,
  [\href{http://arxiv.org/abs/hep-th/0412141}{{\tt hep-th/0412141}}].

\bibitem{Sakai:2005yt}
T.~Sakai and S.~Sugimoto, {\it {More on a holographic dual of QCD}},  {\em
  Prog.Theor.Phys.} {\bf 114} (2005) 1083,
  [\href{http://arxiv.org/abs/hep-th/0507073}{{\tt hep-th/0507073}}].

\bibitem{Atiyah:2004nh}
M.~Atiyah and P.~Sutcliffe, {\it {Skyrmions, instantons, mass and curvature}},
  {\em Phys.Lett.} {\bf B605} (2005) 106,
  [\href{http://arxiv.org/abs/hep-th/0411052}{{\tt hep-th/0411052}}].

\bibitem{Winyard:2015ula}
T.~Winyard, {\it {Hyperbolic Skyrmions}},
  \href{http://arxiv.org/abs/1503.08522}{{\tt arXiv:1503.08522}}.

\bibitem{Bolognesi:2010nb}
S.~Bolognesi and D.~Tong, {\it {Monopoles and holography}},  {\em JHEP} {\bf
  1101} (2011) 153, [\href{http://arxiv.org/abs/1010.4178}{{\tt
  arXiv:1010.4178}}].

\bibitem{Sutcliffe:2011sr}
P.~Sutcliffe, {\it {Monopoles in AdS}},  {\em JHEP} {\bf 1108} (2011) 032,
  [\href{http://arxiv.org/abs/1104.1888}{{\tt arXiv:1104.1888}}].

\bibitem{Piette:1994ug}
B.~Piette, B.~Schroers, and W.~Zakrzewski, {\it {Multisolitons in a
  two-dimensional Skyrme model}},  {\em Z.Phys.} {\bf C65} (1995) 165,
  [\href{http://arxiv.org/abs/hep-th/9406160}{{\tt hep-th/9406160}}].

\bibitem{Bolognesi:2013jba}
S.~Bolognesi and P.~Sutcliffe, {\it {A low-dimensional analogue of holographic
  baryons}},  {\em J.Phys.} {\bf A47} (2014) 135401,
  [\href{http://arxiv.org/abs/1311.2685}{{\tt arXiv:1311.2685}}].

\bibitem{Elliot-Ripley:2015cma}
M.~Elliot-Ripley, {\it {Phases and approximations of baryonic popcorn in a
  low-dimensional analogue of holographic QCD}},  {\em J.Phys.} {\bf A48}
  (2015) 295402, [\href{http://arxiv.org/abs/1503.08755}{{\tt
  arXiv:1503.08755}}].

\bibitem{Salmi:2014hsa}
P.~Salmi and P.~Sutcliffe, {\it {Aloof baby Skyrmions}},  {\em J.Phys.} {\bf
  A48} (2015) 035401, [\href{http://arxiv.org/abs/1409.8176}{{\tt
  arXiv:1409.8176}}].

\bibitem{graham1998dense}
R.~L. Graham, B.~D. Lubachevsky, K.~J. Nurmela, and P.~R. {\"O}sterg{\aa}rd,
  {\it Dense packings of congruent circles in a circle},  {\em Discrete
  Mathematics} {\bf 181} (1998) 139.

\end{thebibliography}\endgroup
\bibliographystyle{JHEP}

\end{document}